\documentclass[preprint,preprintnumbers,amsmath,amssymb,superscriptaddress]{revtex4}
\usepackage{txfonts}% Physical Review B
\usepackage{graphicx}% Include figure files
\usepackage{dcolumn}% Align table columns on decimal point
\usepackage{bm}% bold math
\usepackage{array}
\usepackage{changes}
\usepackage{verbatim}
\usepackage{upgreek}
\usepackage[colorlinks=true,plainpages=false,linkcolor=blue,urlcolor=blue,citecolor=blue,pdfpagemode=UseNone,pdfstartview=FitBH]{hyperref}

\begin{document}

	\title{Antiferromagnetic spin fluctuations and structural transition in cluster Mott insulator candidate Nb$_{3}$Cl$_{8}$ revealed by $^{93}$Nb- and $^{35}$Cl-NMR}
	\author{Y. Z. Zhou}
	\affiliation{Institute of Physics, Chinese Academy of Sciences,\\
		and Beijing National Laboratory for Condensed Matter Physics, Beijing 100190, China}
	\affiliation{School of Physical Sciences, University of Chinese Academy of Sciences, Beijing 100190, China}

   \author{X. Han}
	\affiliation{Institute of Physics, Chinese Academy of Sciences,\\
		and Beijing National Laboratory for Condensed Matter Physics, Beijing 100190, China}
	\affiliation{School of Physical Sciences, University of Chinese Academy of Sciences, Beijing 100190, China}

    \author{J. Luo}
	\affiliation{Institute of Physics, Chinese Academy of Sciences,\\
		and Beijing National Laboratory for Condensed Matter Physics, Beijing 100190, China}

    \author{D. T. Wu}
     \affiliation{Center for Neutron Science and Technology, Guangdong Provincial Key Laboratory of Magnetoelectric Physics and Devices, School of Physics, Sun Yat-Sen University, Guangzhou, Guangdong 510275, China}
     \affiliation{State Key Laboratory of Optoelectronic Materials and Technologies, Sun Yat-Sen University, Guangzhou, Guangdong 510275, China}
     \affiliation{Institute of Physics, Chinese Academy of Sciences,\\
		and Beijing National Laboratory for Condensed Matter Physics, Beijing 100190, China}

     \author{A. F. Fang}
     \affiliation{School of Physics and Astronomy, Beijing Normal University, Beijing 100875, China}
     \affiliation{Key Laboratory of Multiscale Spin Physics, Ministry of Education, Beijing Normal University, Beijing 100875, China}

    \author{B. Shen}
    \affiliation{Center for Neutron Science and Technology, Guangdong Provincial Key Laboratory of Magnetoelectric Physics and Devices, School of Physics, Sun Yat-Sen University, Guangzhou, Guangdong 510275, China}
    \affiliation{State Key Laboratory of Optoelectronic Materials and Technologies, Sun Yat-Sen University, Guangzhou, Guangdong 510275, China}

    \author{B. J. Feng}
	\affiliation{Institute of Physics, Chinese Academy of Sciences,\\
		and Beijing National Laboratory for Condensed Matter Physics, Beijing 100190, China}

    \author{Y. G. Shi}
    \email{ygshi@iphy.ac.cn}
	\affiliation{Institute of Physics, Chinese Academy of Sciences,\\
		and Beijing National Laboratory for Condensed Matter Physics, Beijing 100190, China}
	
    \author{J. Yang}
    \email{yangjie@iphy.ac.cn}
	\affiliation{Institute of Physics, Chinese Academy of Sciences,\\
		and Beijing National Laboratory for Condensed Matter Physics, Beijing 100190, China}

	\author{R. Zhou}
	\email{rzhou@iphy.ac.cn}
	\affiliation{Institute of Physics, Chinese Academy of Sciences,\\
		and Beijing National Laboratory for Condensed Matter Physics, Beijing 100190, China}

	\date{\today}% It is always \today, today,
	
	\begin{abstract}
		{Motivated by recent studies on the cluster mott insulator candidate compound Nb$_{3}$Cl$_{8}$, we perform $^{93}$Nb and $^{35}$Cl nuclear magnetic resonance (NMR) measurements to investigate its electron correlations. Below the structural transition temperature $T_s \sim$ 97 K, all the satellites of the $^{93}$Nb NMR spectra split into three distinct peaks, which suggest a symmetry lowering due to the structural transiton and could be attribute to the  change of the Nb-Nb bond-lengths of Nb$_{3}$ clusters.
%from $P3\overline{m}1$ at low temperatures, either bond length of Nb$_{3}$ clusters changes along the $c$-axis, or the structure of Nb$_{3}$ clusters is no longer equilateral triangle. In addition, we conduct measurements on
The spin-lattice relaxation rate 1/$T_1$ devided by temperature $T$, 1/$T_1$$T$, increases upon cooling down to $T_s$ for all Cl sites, whereas only the Knight shift $K$ of Cl located at the center of Nb$_3$ clusters exhibits a similar temperature dependence as observed in magnetic susceptibility. These findings collectively demonstrate the existence of strong spin correlations between Nb atoms in Nb$_3$Cl$_8$, which should be closely associated with its Mottness.
		}
	\end{abstract}

	%\pacs{74.25.nj, 74.40.-n, 74.25.Dw}
	
	%\keywords{High-temperature superconductors, Nuclear magnetic resonance, pair density wave, Pseudogap}
	
	\maketitle

	In strongly correlated electron system with half-filled bands, significant repulsive Coulomb interaction between electrons leads the system to the % charge localization may induce thus metallic systems with half-filled bands to transform into
so-called Mott insulators\cite{Mott_1968}.  Mott physics has been a perennial hot topic in condensed matter physics, for its appearance and related metal-insulator transition are often intertwined with numerous exotic emergent phenomena, such as the high-temperature superconductivity in doped cuprates\cite{Lee_2006} and quantum spin liquid (QSL)\cite{QSL_2018}. Cluster materials with partially filled bands are also proposed to be Mott insulators, where unpaired electrons are localized within clusters rather than just at lattice sites\cite{Chen_2018}. As for two-dimensional(2D) cluster Mott insulators with kagome lattice, for example, LiZn$_{2}$Mo$_{3}$O$_{8}$ with Mo$_{3}$O$_{13}$ cluster units, the spin states of clusters are affected by electronic correlation and geometric frustration together%could co-act on and modulate spin properties
\cite{Gall_2013,Torardi_1985,Mourigal_2014,Sheckelton_2012}. Topological flat bands resulting from kagome lattice may further complicate the system\cite{Liu_2021,Regnault_2022,Yin_2022}. The interplay between the  spin, charge and orbital degrees of freedom could prompt cluster Mottness to evolve into more exotic states\cite{Sheckelton_2012}, such as charge density wave(CDW)\cite{Sipos_2008,Thomson_1994}, superconductivity\cite{Li_2013} and frustration-driven QSL\cite{Law_2017,Balents_2010} in 1T-TaS$_{2}$.

Recently, there has been growing interests on the 2D van der Waals layered compound Nb$_3$Cl$_8$. The structure of Nb$_3$Cl$_8$ resembles that of Mo$_3$O$_{13}$ cluster-based materials, featuring breathing kagome lattices formed by Nb$_3$Cl$_{13}$ cluster units\cite{Magonov_1993}. At a temperature around $T_s \sim$ 97 K, Nb$_3$Cl$_8$ undergoes a structural phase transition, and the magnetic susceptibility changes from paramagnetic to non-magnetic at low temperatures\cite{Haraguchi_2017}. Previous nuclear magnetic resonance (NMR), X-ray diffraction (XRD) and Terahertz spectroscopic studies report that the structural transition is from the high-temperature (HT) $P3\overline{m}1$ phase to the low-temperature (LT) $R3\overline{m}$ phase\cite{Haraguchi_2017,Pasco_2019,Kim_2023}. NMR measurements suggest charge disproportionation between adjacent layers, which result in the disappearance of unpaired spins below $T_s$\cite{Haraguchi_2017}. However, neutron and high-pressure XRD studies suggest that the space group of the LT phase is monoclinic $C2/m$ accompanied by the C$_3$ symmetry breaking whithin the Nb$_3$ clusters\cite{Sheckelton_2017,Jiang_2022}. Although layer stacking rearrangement is considered to be responsible for the phase transition, the nature of structural transition in  Nb$_3$Cl$_8$ remains controversial.

% due to a second-order Jahn Teller (SOJT) distortion leading to singlet state formation\cite{Sheckelton_2017}. A high-pressure experiment also observed a similar phase transition with $C2/m$ symmetry caused by intra-cluster electron disproportionation\cite{Jiang_2022}. Despite all these studies suggesting layer stacking rearrangement as the quenching mechanism for this phase transition, consensus on its origin has not yet been reached. The nature of such structural transition remains controversial.

Most notably, %In this compound,
each Nb$_3$ trimer in Nb$_3$Cl$_8$ shares an unpaired spin $S_{\rm eff} = 1/2$, which implies a half-filled band at $E_{F}$ and Nb$_3$Cl$_8$ been a potential cluster Mott insulator\cite{Gao_2022}. % and thus suggesting a metallic ground state.
Indeed, electrical conductivity indicates Nb$_3$Cl$_8$ is an insulator\cite{Yoon_2020}.
%However, the insulating behavior and the significant band gap derived from flat bands observed by the angle-resolved photoemission spectroscopy (ARPES) and dynamical mean-field theory (DMFT) calculations\cite{Yoon_2020,Gao_2022}, indicate electron localization.
Density functional theory calculations suggest a half-filled flat band at the Fermi level, whereas angle-resolved photoemission spectroscopy (ARPES) experiments observe a large gap, indicating electron localization and insulating ground state\cite{Gao_2022}.
However, another ARPES study claims a semiconducting ground state with moderate $\sim$ 1.12 eV band gap since inversion symmetry breaking would open up the gap in the Dirac cone\cite{Sun_2022}.
%Therefore, Nb$_3$Cl$_8$ emerges as a candidate for a Mott insulator state based on a single-band Hubbard model.
Recent theories suggest that the discrepancy of different works could be attributed to band-selective normalization, and  Nb$_3$Cl$_8$ is a cluster Mott insulator described by single-band Hubbard model\cite{Hu_2023,Grytsiuk_2024}.
% The half-filled flat band near EF splits and forms a Mott gap, while the observed flat band by ARPES near $E_{F}$ may correspond to the lower Hubbard band\cite{Hu_2023}.
Although magnetic susceptibility exhibits a Curie-Weiss behavior at high temperatures and negative Weiss temperature indicates antiferromagnetic interactions between Nb$_3$ trimers\cite{Haraguchi_2017}, microscopic investigation on spin correlations in Nb$_3$Cl$_8$ is still lacking. There remains ongoing debate regarding the ground state and electron correlations of  Nb$_3$Cl$_8$.
 %Nb$_3$Cl$_8$ at elevated temperatures to determine whether intrinsic Curie-Weiss behavior is present in its magnetic susceptibility. Henceforth, there remains ongoing debate regarding whether Nb$_3$Cl$_8$ is indeed a Mott insulator containing electron correlations.
The clarification of this issue holds particular significance, given that numerous theoretical studies predict novel properties such as possible QSL\cite{Balents_2010,Hu_2023,Kasahara_2018}, monolayer ferromagnetism\cite{Jiang_2017}, topological insulation\cite{Bradlyn_2017}, unconventional superconductivity through doping of the Mott state\cite{Zhang_2023}, and even potential novel properties in heterostructures\cite{Wu_2022,Ando_2020}. %Further advancements in microscopic research techniques are imperative for investigating electron correlations in Nb$_3$Cl$_8$.

%
	%The controversies mentioned above and our curiosity about the strong correlation types related to Mott physics in Nb$_{3}$Cl$_{8}$ motivate us to conduct microscopic research methods, such as NMR, for more in-depth investigation.
	
	In this study, we conducted both $^{93}$Nb- and $^{35}$Cl-NMR measurements to investigate electron correlations and local structure in Nb$_{3}$Cl$_{8}$. In the LT phase below $T_s$, the satellite lines of $^{93}$Nb-NMR spectra split into  three lines with nearly identical spectral weight, indicating there exists three inequivalent Nb sites and the bond-lengths of Nb$_3$ clusters change due to structural transition.
%the presence of three distinct Nb sites. Our findings indicate that  Nb$_{3}$ clusters become unequivalent through bond length variations.
In the HT phase, the spin-lattice relaxation rate 1/$T_1$ devided by temperature $T$, 1/$T_1$$T$, shows a monotonic increase for all Cl sites as temperature decreases towards $T_s$. However, only the Knight shift of Cl located at the center of the Nb$_3$ cluster displays a similar temperature dependence as observed in magnetic susceptibility, implying that Nb$_3$ trimer indeed shares an unpaired spin $S_{\rm eff} = 1/2$. These results provide microscopic evidence for antiferromagnetic spin fluctuations at high temperatures in Nb$_{3}$Cl$_{8}$ and suggest the existence of Mott physics. Furthermore, our findings highlight how geometric frustration influences electronic correlations through its unique cluster kagome structure and offer new insights into its phase transition and ground state.
	
The Nb$_{3}$Cl$_{8}$ single crystals utilized in this study were synthesized via the PbCl$_{2}$ flux method\cite{Haraguchi_2017}. The typical dimensions of a single crystal are 2mm$\times$2mm$\times$0.1mm.
The $^{93}$Nb- and $^{35}$Cl-NMR spectra were obtained through fast Fourier transform (FFT) summation. The spin echo was observed by using a $\pi$/2-$\tau$-$\pi$ pulse sequence on a phase coherent spectrometer. A $\pi$/2 = 5 $\mu$s pulse was employed with pulse intervals $\tau$ = 150 $\mu$s for $^{35}$Cl and 75 $\mu$s for $^{93}$Nb nuclei. The spin-lattice relaxation time ($T_1$) was determined using a saturation recovery method.	

%Based on the previous NMR research of Nb$_{3}$Cl$_{8}$\cite{Haraguchi_2017}, our measurement is implemented to the $^{93}$Nb nuclei in the LT phase at first.

	\begin{figure}[htbp]
		\includegraphics[width= 15 cm]{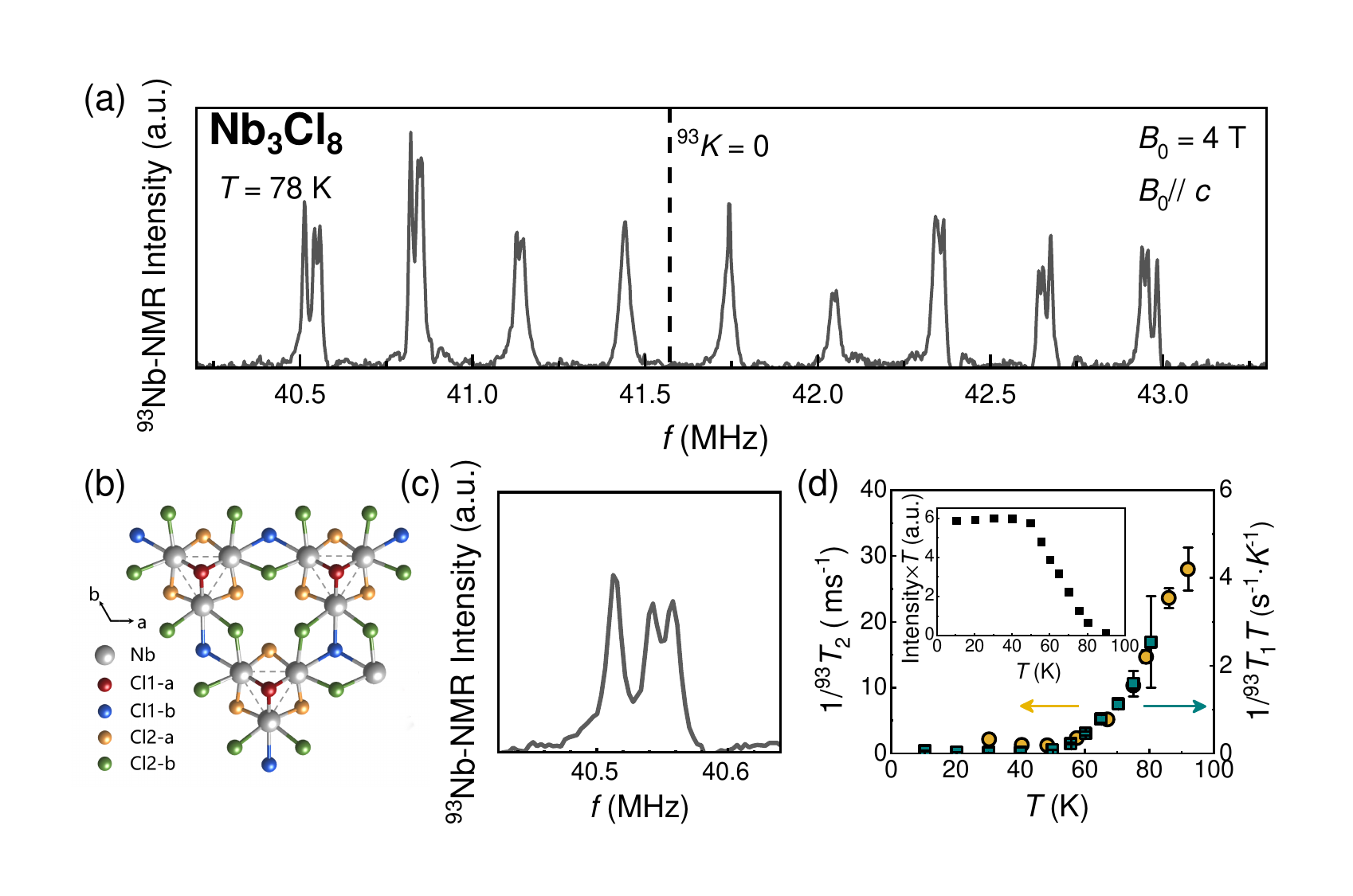}
		\caption{(Color online) (a) The $^{93}$Nb NMR spectrum at 78 K with the applied magnetic field $B_{0}$ = 4 T. The dashed line shows Knight shift $K$ = 0. (b) Kagome lattice structure of monolayer Nb$_{3}$Cl$_{8}$. (c) The enlarged view of the lowest frequcncy satellite peak of the $^{93}$Nb NMR spectrum. (d) The temperature variation of 1/$T_1T$(square), 1/$T_{2}$(circle) and spectrum weight(inset) of $^{93}$Nb.
			\label{Nb_spec}}
	\end{figure}

Figure \ref{Nb_spec}(a) shows the $^{93}$Nb NMR spectrum at 78 K under a fixed magnetic field $B_{0}$ = 4 T along the $c$-axis. For nucleus with the nuclear quadrupole moment $Q$, the total Hamiltonian under field can be written as:
	\begin{equation}
		\label{eq1}
		\mathcal{H} = \mathcal{H}_{0} + \mathcal{H}_{\rm Q} = \gamma\hbar\textbf{\emph{I}}\cdot\textbf{\emph{B}}_{0}(1+K)+\dfrac{e^{2}qQ}{4I(I+1)}[(3I^{2}_{z}-I^{2})+\dfrac{1}{2}\eta(I^{2}_{+}+I^{2}_{-})]
	\end{equation}		
	where $K$ is the Knight shift, $^{93}\gamma$ = 10.405 MHz/T is the nuclear gyromagnetic ratio , $eq$ is the electric field gradient (EFG) along the principal axis and $\eta = |\tfrac{V_{xx} - V_{yy}}{V_{xx} + V_{yy}}|$ is the EFG asymmetry parameter. Since $^{93}$Nb nucleus has a spin $I$ = 9/2, the NMR spectrum is expected to exhibit nine resonance peaks, i.e., one central peak and eight first-order satellites. In the HT phase above $T_s$, $^{93}$Nb has only one crystallographic site, as shown in Fig.\ref{Nb_spec} (b). However, at $T$ = 78 K that is below $T_s$, all satellite peaks split into three peaks except for the central line. Figure \ref{Nb_spec}(c) shows an enlarged view of the lowest frequency satellite peak. The area ratio of the three peaks is approximately 1:1:1, indicating the presence of three distinct Nb sites in the LT phase. Table 1 shows the obtained nuclear quadrupole resonance (NQR) frequency $\nu _Q$, $K$, $\eta$ and relative abundance ratio (AO) for the three $^{93}$Nb sites from the $^{93}$Nb NMR spectrum.
%Our findings strongly suggest that there exists a discernible difference between the crystal structures of the LT and high-temperature (HT) phases, as illustrated in Fig. 2(b), where only one Nb site is present. However,
It is noted that in previous NMR measurements only two set of split peaks with an area ratio close to 1:1 were observed, which was interpreted as two alternating nonmagnetic layers of Nb$_3$ clusters due to charge disproportionation\cite{Haraguchi_2017} .
In our study, we employ a much smaller single crystal sample and apply a magnetic field strength of only 4 T, resulting in narrower NMR lines. Considering that the two higher frequency peaks are closely spaced as shown in Fig. \ref{Nb_spec}(c), they would be hardly resolved using larger samples at higher fields.

	\begin{table}
	\centering
	\caption{NMR parameters for $^{93}$Nb spectrum at 78 K}
	\begin{ruledtabular}
		\begin{tabular}{ccccc}
		site & $\nu_{Q}$ & $K(\%)$ & $\eta$ & AO \\
		\hline
		Nb1 & 0.617 & 0.44 & 0.019 & $\sim$1/3 \\
		Nb2-1 & 0.603 & 0.44 & 0.027 & $\sim$1/3 \\
		Nb2-2 & 0.596 & 0.44 & 0.027 & $\sim$1/3 \\
		\end{tabular}
	\end{ruledtabular}
	\end{table}

	\begin{figure}[htbp]
	\includegraphics[width= 14 cm]{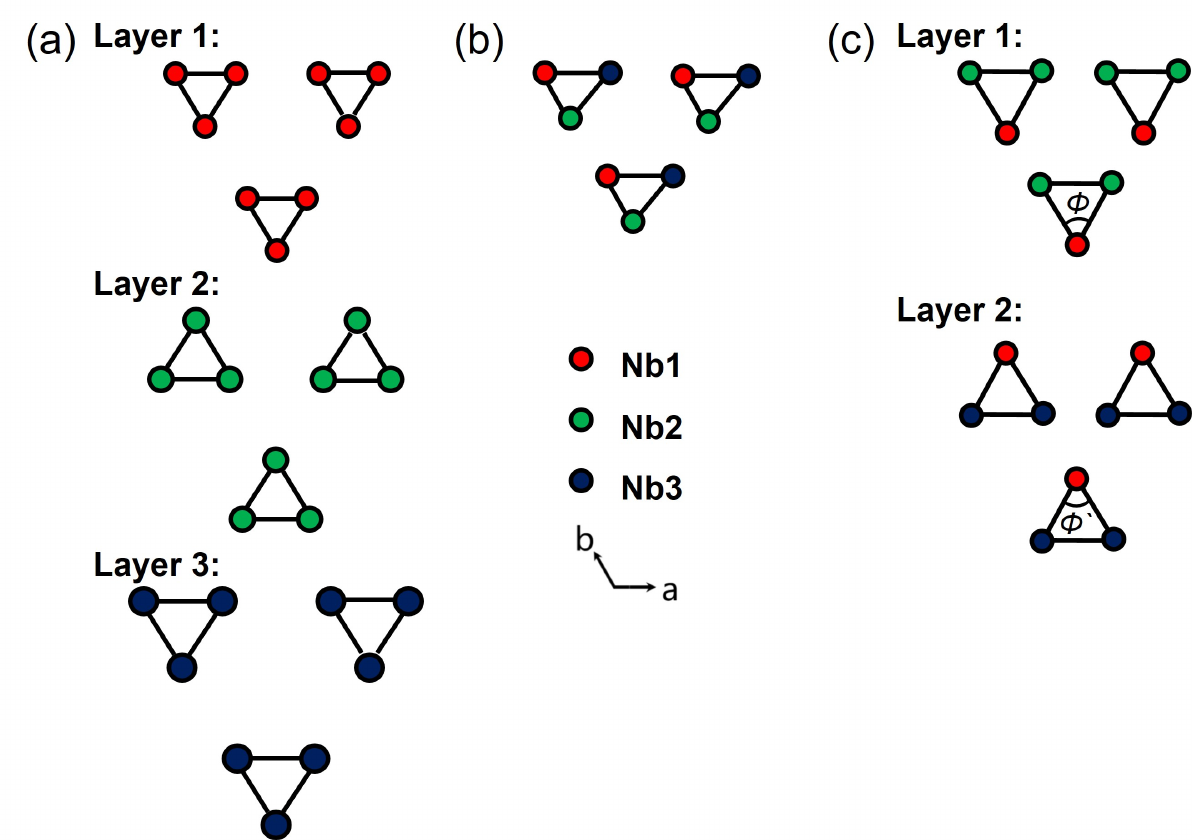}
	\caption{(Color online) The sketches of possible Nb$_{3}$ trimer structures within adjacent layers in the LT phase. (a) The Nb$_3$ trimers remain equilateral triangles, but have different Nb-Nb bond-lengths for adjacent layers, resulting in a 3$c$ periodicity along $c$-axis. Notably, this configuration does not break rotational symmetry. (b) Distinct Nb-Nb bond-lengths for Nb$_3$ trimer within one layer.
 %while maintaining identical layer structures.
(c) Nb$_3$ trimer has an isosceles triangle configuration in each layer but with different top angles $\phi$ for adjacent layers, leading to a 2$c$ period along $c$-axis. It is noted that both (b) and (c) result in  rotational symmetry breaking. For all cases, three different Nb sites should be observed in NMR spectra.
		\label{Sketch}}
	\end{figure}

Next we discuss the reason for the presence of three distinct Nb positions in the LT phase. One explanation is that the Nb$_3$ cluster maintains the same equilateral triangular structure as in the HT phase, but Nb-Nb bond-lengths vary along the $c$ direction as shown in Fig. \ref{Sketch}(a). Alternatively, it is plausible that the Nb$_3$ clusters deviate from  equilateral triangle, resulting in three inequivalent Nb sites as shown in Fig. \ref{Sketch}(b).
We note that neutron scattering and XRD studies suggest a monoclinic $C2/m$ space group for the LT phase and an isosceles triangle structure for the Nb$_3$ cluster\cite{Sheckelton_2017}. Such  configuration of Nb$_3$ clusters would result in only  two different Nb sites with an atomic ratio of 1:2. Therefore, another possibility is that  Nb$_3$ clusters form an isosceles triangle structure along the $c$ axis but with different  top angles $\phi$ for adjacent layers as shown in Fig. \ref{Sketch}(c). To fully determine the LT structure of Nb$_{3}$Cl$_{8}$, further high-resolution XRD studies are still required. In addition, we observe that the central lines of the three Nb sites coincide as shown in Fig. \ref{Nb_spec}(a), indicating they have identical Knight shift. This is consistent with the fact that magnetic susceptibility in the LT phase is nearly zero, which causes no spin shift at any of these Nb sites. Our results reveal the appearance of inequivalent Nb$_3$ clusters in the LT phase, which disrupts shared unpaired spins within overlapping molecular orbitals and consequently leads to non-magnetic behavior.
%within overlapping molecular orbitals

%This explanation could be further validated by the two splitting satellite peaks with atomic ratio approximately around 1:2 at 10 K, as shown in supplymentary\cite{SM}.

%We consider that the three-peak-splittings at 78 K can be attributed to environmental differences in two Nb nuclei with shorter bond length, because the frequencies of the outermost splitting satellite peaks hardly change with temperature, while the frequencies of the other two satellite peaks are closer to each other. In this way, we prefer to regard the phase transition as a result of the threefold rotational symmetry breaking from trigonal $P3\overline{m}1$ to monoclinic $C2/m$.

With increasing temperature, the $^{93}$Nb-NMR signal intensity decreases significantly due to a sharp increase in the spin-spin relaxation rate 1/$T_{2}$ and  $1/T_{1}T$ near the structural phase transition as shown in Fig. \ref{Nb_spec}(d). These observations strongly suggest a rapid enhancement of spin fluctuations near $T_s$ with increasing temperature. Since  $^{93}$$T_{2}$ is much shorter compared to the dead time of our spectrometer ($\sim$5us), the $^{93}$Nb-NMR signal is completely wiped out above $T_s$. Therefore, to gain insight into the  electron correlations in the HT phase, we then performed $^{35}$Cl-NMR measurements. All Cl sites are far from Nb atoms, thus the hyperfine coupling between $^{35}$Cl nuclei and Nb electron spins should be significantly weaker, which results in a much longer spin-spin relaxation time for $^{35}$Cl-NMR signals and ensures their persistence in the HT phase.%so HT phase research can be conducted by $^{35}$Cl nuclei.

	\begin{figure}[htbp]
		\includegraphics[width= 13 cm]{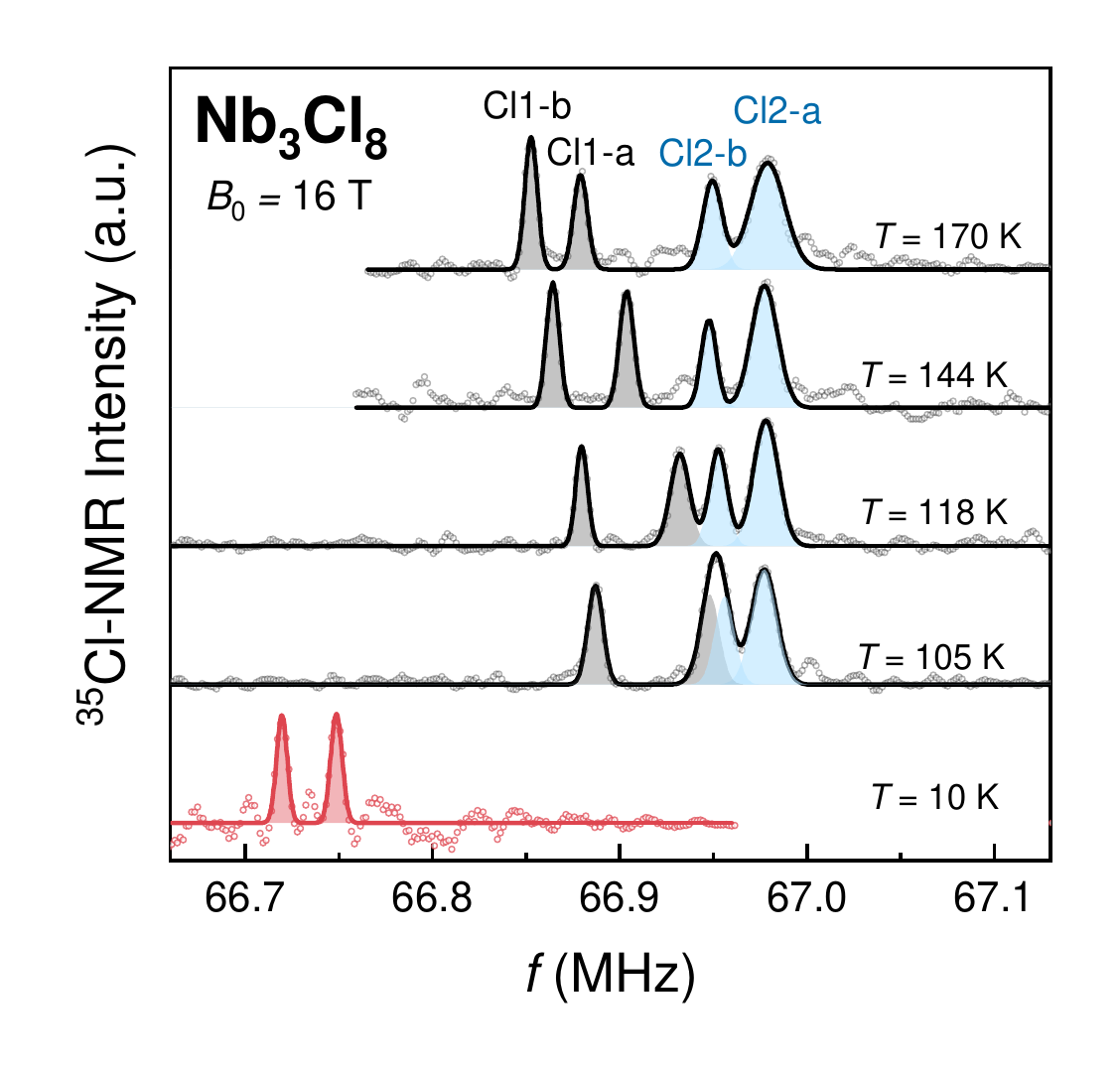}
		\caption{(Color online)  The temperature dependence of the $^{35}$Cl-NMR spectra under an applied magnetic field $B_0$ = 16 T for Nb$_{3}$Cl$_{8}$ along the $c$ axis. The solid curves and shaded regions are the fittings for different Cl sites by Gaussian function.
			\label{Cl_spec}}
	\end{figure}

Figure \ref{Cl_spec} shows the $^{35}$Cl-NMR spectra at different temperatures under an applied magnetic field $B_0$ = 16 T along the $c$-axis. In Nb$_3$Cl$_8$, there are two different Cl crystallographic sites, %as depicted in Fig.3.
namely, Cl1 in the center of the Nb-triangle and Cl2 beside the Nb-triangle, as shown in Fig. \ref{Nb_spec}(b). As a result of the different Nb-Nb bond-lengths within the layer, both Cl1 and Cl2 sites can further be divided into two types: Cl1-a and Cl2-a near the shorter Nb-Nb bond length, and Cl1-b and Cl2-b near the longer Nb-Nb bond length. For $^{35}$Cl nucleus with a spin of $I$ = 3/2, each Cl site should exhibit three transition lines in the NMR spectrum, thus the spectrum should have twelve lines overall. However, due to limitations imposed by single crystal size, no satellite peaks were observed and instead only four central lines were visible above $T_s$. In the following, we elaborate how the $^{35}$Cl sites are identified in the NMR spectra. We measured the magnetic field dependence of the $^{35}$Cl-NMR spectra and obtained the EFG asymmetry parameter $\eta$  for each Cl peak.
%To identify these NMR lines, we investigated their resonance frequency dependence versus magnetic fields  and found that
The two lower frequency peaks exhibit nearly zero asymmetry parameter $\eta$, whereas two higher frequency peaks display finite $\eta$ values (further details and additional results can be found in supplemental materials\cite{SM}). Since Cl1-a and Cl1-b locate in the symmetric EFG enviroment,  $\eta$ for Cl1-a and Cl1-b sites should be zero. Therefore, we assign
%The isotropic nature of in-plane electric field gradient (EFG) at both Cl1-a and Cl1-b sites renders $\eta$ equal to zero. However, for the adjacent locations of Cl2-a and Cl2-b next to Nb-Nb bonds where EFG becomes anisotropic, $\eta$ deviates from zero value accordingly. Consequently, we can assign
the two lower frequency peaks as corresponding to Cl1 sites, while the two higher frequency peaks to Cl2 sites. The Cl-a and Cl-b
can be further distinguished by comparing their hyperfine coupling constants, as described below. Upon decreasing temperature below $T_s$, an abrupt change of the  NMR spectra is observed, indicating a first-order structural transition. Only two peaks corresponding to Cl1-a and Cl1-b sites can be observed at 10 K. A possible reason for the change of the spectra is that Cl2 sites have longer $T_1$, resulting in an undetectable NMR signal for Cl2.

%Finally, the independent variation trends of the four peaks confirm the statement that they belong to four different sites in Nb$_{3}$Cl$_{13}$ unit cell. The peak $\rm \uppercase\expandafter{\romannumeral1}$ and $\rm \uppercase\expandafter{\romannumeral2}$ points to the their origin from isotropic sites, that is, the Cl1-b and C1-a sites in the cental position of three Nbs.

%Comparing their value of $A_{\rm hf}$ calculated by $K-\chi$ plot, the C1-a sites could be further confirmed as the Cl inside Nb$_{3}$ cluster.

%While the finite value of peak $\rm \uppercase\expandafter{\romannumeral3}$ and $\rm \uppercase\expandafter{\romannumeral4}$ corresponds to anisotropic Cl2-b and Cl2-a sites. The structural relationships of Cl1 and Cl2 between Nb$_{3}$ cluster are shown in Fig.1(b) and Fig.3.

The Knight shift $K$ is defined as ${{K}}=\left( f-{{f}_{2\text{nd}}}-\gamma {{B}_{0}} \right)/\gamma {{B}_{0}}$, where $f$ is the observed resonance frequency, $\gamma$ = 4.17169MHz/T is the gyromagnetic ratio of $^{35}$Cl, and ${f}_{2\text{nd}}$ is the second-order quadrupole shift (more details can be seen in the supplemental materials\cite{SM}). Figure \ref{Cl_K&T1}(a) shows the temperature dependence of $K$ for all Cl sites. The most striking feature
is that $K$ for Cl1 and Cl2 show completely different variation. For Cl1-a and Cl1-b sites, $K$ presents a Curie-Weiss behavior, which is consist with the temperature variation of magnetic susceptibility $\chi$\cite{Haraguchi_2017}.
The $K$ can be written as ${{K}}=K_{\text{spin}}+K_{\text{orb}}$.
%=A^{\text{spin}}\chi _{cc}^{\text{spin}}+A^{\text{orb}}\chi _{cc}^{\text{orb}}$,
%\begin{equation}
%\label{K1}
%\begin{aligned}
%{{K}_{c}}=K_{c}^{\text{spin}}+K_{c}^{\text{orb}}=A_{c}^{\text{spin}}\chi _{cc}^{\text{spin}}+A_{c}^{\text{orb}}\chi _{cc}^{\text{orb}}
%\end{aligned}
%\end{equation}
Here, $K_{\text{spin}} = A_{\text{spin}}\chi _{\text{spin}}$ is proportional to the static spin susceptibility $\chi_{\text{spin}}$ by hyperfine coupling constant $A_{\text{spin}}$, while $K_{\text{orb}}$ is the contribution from orbital (Van Vleck) susceptibility and usually temperature independent.
%and $\chi_{cc}^{\text{orb}}$ are the spin and orbital(Van Vleck) susceptibility, respectively. $A_{c}^{\text{spin}}$ and $A_{c}^{\text{orb}}$ are the hyperfine coupling constants related to the spin and orbital susceptibility, respectively.
By plotting $K$ as a function of $\chi$, we find that they hold linear relation above $T_s$\cite{SM}, indicating the intrinsic Curie-Weiss behavior of $\chi_{\text{spin}}$. The $A_{\text{spin}}$ values for Cl1-a and Cl1-b are different, which allow us to discriminate between the two Cl sites. Since  Cl1 is located in the center of Nb$_3$ triangle, our findings provide clear microscopic evidence for the presence of unpaired spins on the molecular orbitals of each Nb$_3$ trimer. Through fitting $K$ with the Curie-Weiss function as shown in Fig. \ref{Cl_K&T1}(a), we obtain the Weiss temperature $\theta = -20 \pm 10$ K. The negative value of $\theta$ suggests antiferromagnetic interactions between Nb$_3$ trimers.
For Cl2-a and Cl2-b sites, $K$ is nearly temperature-independent. This might be because that the hyperfine coupling constant happens to be almost zero for Cl2 sites~\cite{SM}.

\begin{figure}[htbp]
		\includegraphics[width= 11 cm]{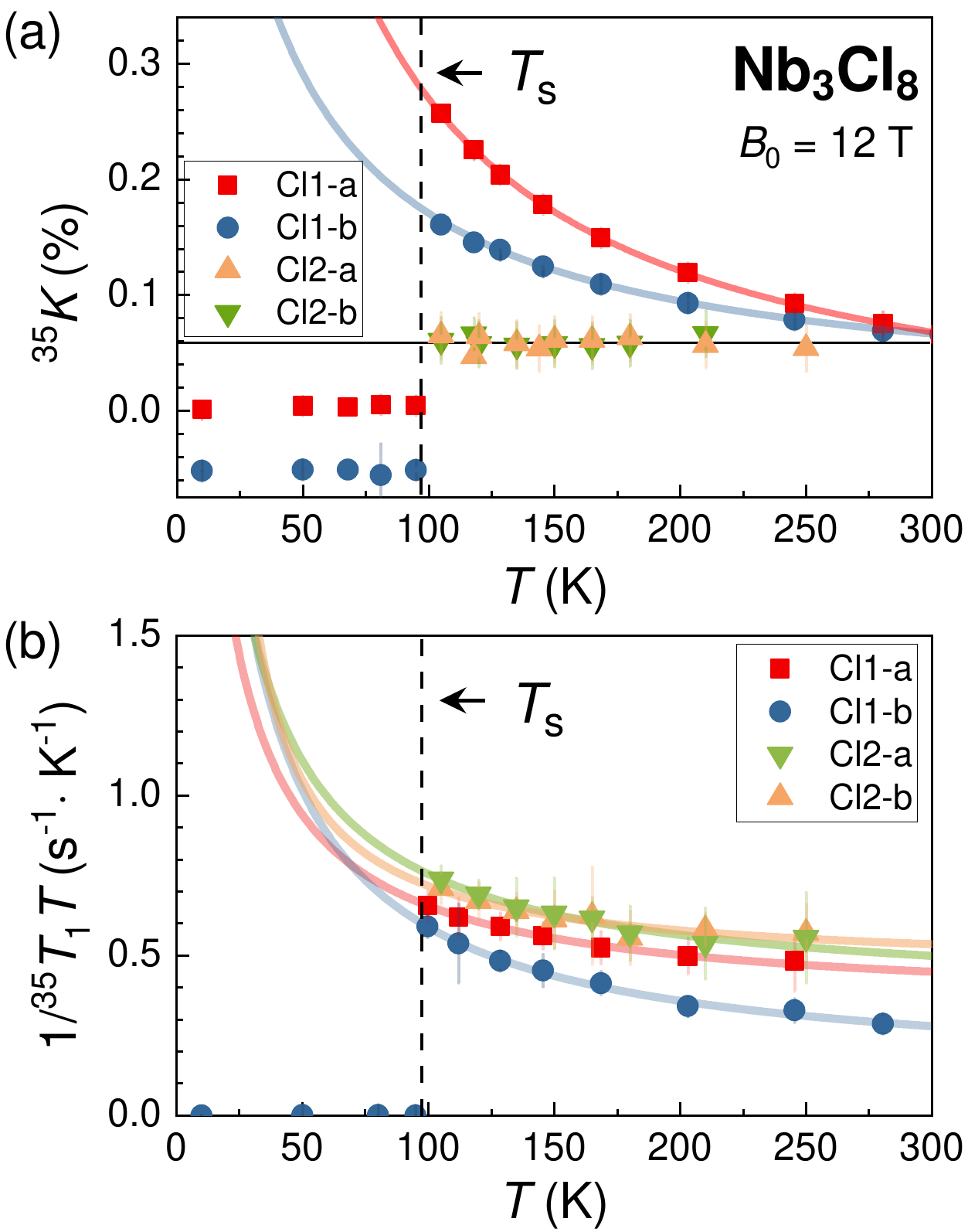}
		\caption{(Color online) The temperature dependence of  $K$(a) and 1/$T_1T$(b) for different $^{35}$Cl sites. The dashed line indicates the structural phase transition temperature $T_{\rm s}$. The solid lines are Curie-Weiss fits.
		\label{Cl_K&T1}}
	\end{figure}

Last, we discuss the spin correlations based on the temperature dependence of $1/T_{1}T$ as shown in Fig. \ref{Cl_K&T1}(b).  For all Cl sites, $1/T_{1}T$  increases with decreasing temperature until reaching $T_s$, followed by a sudden transition to an extremely low value below $T_s$. Notably, this value is approximately 430 times smaller than that observed for $^{93}$Nb site (Fig. \ref{Nb_spec}(d)). Consequently, the $^{93}$Nb-NMR signal in the HT phase would be wiped out due to the pronounced fluctuations of Nb 4$d$ electron spins. All these findings strongly support the presence of spin fluctuations in the HT phase and provide compelling evidence for Nb$_3$Cl$_8$ being a cluster Mott insulator. In general,
$1/T_{1}T$ is proportional to the imaginary part of the dynamic
susceptibility perpendicular to the applied field ($\chi''_\perp(q,\omega)$) and can be written as
\begin{equation}
\frac{1}{T_1T} \propto \sum_q |A(\textbf{q})|^2 \frac{\chi''_\perp(\textbf{q},\omega)}{\omega},
\end{equation}
where $A(\textbf{q})$ is the wave
vector \textbf{q} dependent hyperfine coupling constant and $\omega$ is the NMR frequency.
The $K$ probes the static magnetic susceptibility at \textbf{q} = 0, while $1/T_{1}$ is also sensitive to the spin fluctuations at
$\textbf{q} \neq 0$. There will be a peak around a finite wave vector (due to antiferromagnetic spin fluctuations). Therefore, $A(\textbf{q})$ in $1/T_{1}T$ is significant for all Cl sites, though $A$ happens to be nearly zero for Cl2 sites in the Knight shift, as that observed in YBa$_{2}$Cu$_{3}$O$_{6.63}$ \cite{takigawa}.
To gain insight into the spin fluctuations, 1/$T_1T$ can be written as ${\dfrac{1}{T_1T}} = {\left (\dfrac{1}{T_1T}\right )_{0}} + {\left (\dfrac{1}{T_1T}\right )_{\rm AF}}$. The first term $(1/T_1T)$$_{0}$ represents the contribution from the non-correlation electrons and is related to the density of states at the Fermi level. The second term $(1/T_1T)_{\rm AF}$ represents the contribution from the antiferromagnetic spin fluctuations. In the case of 2D antiferromagnetic spin fluctuations, $(1/T_1T)_{\rm AF}$ follows a Curie-Weiss $T$ dependence as\cite{Moriya},
	\begin{equation}
		\label{eq2}
		(1/T_1T)_{\rm AF}\propto{\dfrac{C}{T+\theta}}
	\end{equation}
	%where $\theta$ is a symbolic temperature related to AF correlations.
As shown in Fig. \ref{Cl_K&T1}(b), the experimental data can be well-fitted by Eq.(3). The obtained $\theta \sim - 30 \pm 20$ K suggests that the antiferromagnetic spin fluctuations would persistently be enhanced if the structural transition does not take place at $T_s \sim$  97 K. The $1/T_{1}T$ results also imply that the structural transition is not triggered by the antiferromagnetic spin correlations. However, it should be noted that no signature of a magnetic transition was observed in the powder sample, in which the structural transition is suppressed\cite{Haraguchi_2017}. Therefore, our results indicate that powder Nb$_3$Cl$_8$ could be a QSL candidate due to its strong electron spin correlations and kagome structure. Further investigations are still required to fully elucidate the ground state properties of powder Nb$_3$Cl$_8$.

In summary, we investigate the electron correlations and local structure in Nb$_{3}$Cl$_{8}$ by $^{93}$Nb- and $^{35}$Cl-NMR measurements. In the LT phase, we observed that the satellite lines of the $^{93}$Nb-NMR spectra exhibit three lines with nearly identical spectral weight, indicating the presence of three distinct Nb sites. Our findings suggest that bond length variations result in the inequivalence among Nb$_{3}$ clusters. In the HT phase, 1/$T_1$$T$ shows a Curie-Weiss temperature dependence for all Cl sites as temperature decreases towards $T_s$. The Knight shift of Cl located at the center of the Nb$_3$ cluster also shows a similar temperature dependence as observed in magnetic susceptibility, implying an unpaired spin shared by the Nb$_3$ trimer. Thus, our results indicate that antiferromagnetic spin fluctuations develop at high temperatures in Nb$_{3}$Cl$_{8}$ . We anticipate that our work will inspire further experimental measurements on Nb$_{3}$Cl$_{8}$ to search for novel phenomena and related Mott physics.

\begin{acknowledgments}
%	The authors thank  for valuable discussions.
This work was supported by National Key Research and Development Projects of China (Grant No. 2022YFA1403402, No. 2023YFA1406103, No. 2024YFA1409200, No. 2024YFA1611302 and No. 2023YFF0718400), the National Natural Science Foundation of China (Grant No. 12374142, No. 12304170, W2411004 and 12374197), the Beijing Natural Science Foundation (JQ23001), the Strategic Priority Research Program of the Chinese Academy of Sciences (Grant No.XDB33010100) and the Guangdong basic and applied basic research foundation (2023B151520013). This work was supported by the Synergetic Extreme Condition User Facility (SECUF, https://cstr.cn/31123.02.SECUF).
\end{acknowledgments}


\begin{references}
\itemsep=-1pt plus.2pt minus.2pt
	

\bibitem{Mott_1968}
Mott N F 1968 \emph{Rev. Mod. Phys.} {\bf 40} 677


\bibitem{Lee_2006}
Lee P A, Nagaosa N, and Wen X G 2006 {\it Rev. Mod. Phys.} {\bf 78} 17


\bibitem {QSL_2018}
Pustogow A, Bories M, L\"{o}hle A, R\"{o}sslhuber R, Zhukova E, Gorshunov B, Tomi\'{c} S, Schlueter J A, H\"{u}bner R, Hiramatsu T, Yoshida Y, Saito G, Kato R, Lee T, Dobrosavljevi\'{c} V, Fratini S, and  Dressel M 2018 {\it Nature Mater.} {\bf 17} 773


\bibitem {Chen_2018} Chen G, Lee P A 2018 {\it Phys. Rev. B} {\bf 97} 035124

	
\bibitem {Gall_2013} Gall P, Al Rahal Al Orabi R, Guizouarn T, and Gougeon P 2013 {\it J. Solid State Chem.} {\bf 208} 99
	
	
\bibitem {Torardi_1985} Torardi C C and McCarley R E 1985 {\it Inorg. Chem.} {\bf 24} 476
	

\bibitem {Mourigal_2014} Mourigal M, Fuhrman W T, Sheckelton J P, Wartelle A, Rodriguez-Rivera J A, Abernathy D L, McQueen T M, and Broholm C L 2014 {\it Phys. Rev. Lett.} {\bf 112} 027202
	
	
\bibitem {Sheckelton_2012} Sheckelton J P, Neilson J R, Soltan D G, and McQueen T M 2012 {\it Nat. Mater.} {\bf 11} 493
	
	
\bibitem {Liu_2021} Liu H, Meng S, and Liu F 2021 {\it Phys. Rev. Mater.} {\bf 5} 084203
	

\bibitem {Regnault_2022} Regnault N, Xu Y, Li M R, Ma D S, Jovanovic M, Yazdani A, Parkin S S P, Felser C, Schoop L M, Ong N P, Cava R J, Elcoro L, Song Z D, and Bernevig B A 2022 {\it Nature (London)} {\bf 603} 824
	


\bibitem {Yin_2022} Yin I X, Lian B, and Hasan M Z 2022 {\it Nature (London)} {\bf 612} 647
	

%	\bibitem{Kang_2020}
%	M. Kang, S. Fang, L. Ye, H. C. Po, J. Denlinger, C. Jozwiak, A. Bostwick, E. Rotenberg, E. Kaxiras, and J. G. Checkelsky et al., Topological flat bands in frustrated kagome lattice CoSn, \href{https://doi.org/10.1038/s41467-020-17465-1} {\emph{Nat. Commun.} {\bf 11}, 4004 (2020).}
	


\bibitem {Sipos_2008} Sipos B, Kusmartseva A F, Akrap A, Berger H, Forro L, and Tutis E 2008 {\it Nat. Mater.} {\bf 7} 960
	
	%\bibitem{Sipos_2008}
	%B. Sipos, A. F. Kusmartseva, A. Akrap, H. Berger, L. Forro, and E. Tutis, From Mott state to superconductivity in 1$T$-TaS$_{2}$,
%\href{https://doi.org/10.1038/nmat2318} {\emph{Nat. Mater.} {\bf 7}, 960 (2008).}
	
	
\bibitem {Thomson_1994} Thomson R E, Burk B, Zettl A, and Clarke J 1994 {\it Phys. Rev. B} {\bf 49} 16899

	%\bibitem{Thomson_1994}
	%R. E. Thomson, B. Burk, A. Zettl, and John Clarke, Scanning tunneling microscopy of the charge-density-wave structure in 1$T$-TaS$_{2}$, \href{https://link.aps.org/doi/10.1103/PhysRevB.49.16899} {\emph{Phys. Rev. B} \textbf{49}, 16899 (1994).}
	
\bibitem {Li_2013} Li L J, Lu W J, Liu Y, Qu Z, Ling L S, and Sun Y P 2013 {\it Physica C} {\bf 492} 64

	%\bibitem{Li_2013}
	%L. J. Li, W. J. Lu, Y. Liu, Z. Qu, L.S. Ling, and Y. P. Sun, Influence of defects on charge-density-wave and superconductivity in 1$T$-TaS$_{2}$ and 2$H$-TaS$_{2}$ systems \href{https://doi.org/10.1016/j.physc.2013.06.002} {\emph{Physica C} {\bf 492}, 64 (2013).}
	
\bibitem {Law_2017} Law K T, and Lee P A 2017 {\it Proc. Natl. Acad. Sci.} {\bf 114} 6996

	%\bibitem{Law_2017}
	%K. T. Law, and P. A. Lee, 1$T$-TaS$_{2}$ as a quantum spin liquid, \href{https://doi.org/10.1073/pnas.1706769114} {\emph{Proc. Natl. Acad. Sci. } {\bf 114}, 6996 (2017).}


\bibitem {Balents_2010} Balents L 2010 {\it Nature} {\bf 464} 199
	
	%\bibitem{Balents_2010}
	%L. Balents, Spin liquids in frustrated magnets, \href{https://www.nature.com/articles/nature08917}{\emph{Nature} \textbf{464}, 199 (2010).}


\bibitem {Magonov_1993} Magonov S N, Zoennchen P, Rotter H, Cantow H J, Thiele G, Ren J, and Whangbo M H 1993 {\it J. Am. Chem. Soc.} {\bf 115} 2495
	
	
	%\bibitem{Magonov_1993}
	%S. N. Magonov, P. Zoennchen, H. Rotter, H. J. Cantow, G. Thiele, J. Ren, and M. H. Whangbo, Scanning tunneling and atomic force microscopy study of layered transition metal halides Nb$_ {3}X_ {8}$ ($X$ = Cl, I), \href{https://doi.org/10.1021/ja00059a053}{\emph{J. Am. Chem. Soc.} \textbf{115}, 2495 (1993).}


\bibitem {Haraguchi_2017} Haraguchi Y, Michioka C, Ishikawa M, Nakano Y, Yamochi H, Ueda H, and Yoshimura K 2017 {\it Inorg. Chem.} {\bf 56} 3483
	
	
	%\bibitem{Haraguchi_2017}
	%Y. Haraguchi, C. Michioka, M. Ishikawa, Y. Nakano, H. Yamochi, H. Ueda, and K. Yoshimura, Magnetic-Nonmagnetic Phase Transition with Interlayer Charge Disproportionation of Nb$_{3}$ Trimers in the Cluster Compound Nb$_ {3} $Cl$_ {8}$, \href{https://pubs.acs.org/doi/10.1021/acs.inorgchem.6b03028} {\emph{Inorg. Chem.}  \textbf{56}, 3483 (2017).}
	
\bibitem {Pasco_2019} Pasco C M, El Baggari I, Bianco E, Kourkoutis L F, and McQueen T M 2019 {\it ACS Nano} {\bf 13} 9457
	
	%\bibitem{Pasco_2019}
	%C. M. Pasco, I. El Baggari, E. Bianco, L. F. Kourkoutis, and T. M. McQueen, Tunable Magnetic Transition to a Singlet Ground State in a 2D van der Waals Layered Trimerized Kagom¨¦ Magnet, \href{https://doi.org/10.1021/acsnano.9b04392}{\emph{ACS Nano} {\bf 13}, 9457 (2019)}.

\bibitem {Kim_2023} Kim J, Lee Y, Choi Y W, Jung T S, Son S, Kim J, Choi H J, Park J -G, and Kim J H 2023 {\it ACS Omega} {\bf 8} 14190


%\bibitem{Kim_2023}
	%J. Kim, Y. Lee, Y. W. Choi, T. S. Jung, S. Son, J. Kim, H. J. Choi, J.-G. Park, and J. H. Kim, Terahertz Spectroscopy and DFT Analysis of Phonon Dynamics of the Layered Van der Waals Semiconductor Nb$_ {3}X_ {8}$ ($X$ = Cl, I), \href{https://doi.org/10.1021/acsomega.3c01019}{\emph{ACS Omega} {\bf 8}, 14190 (2023)}.


\bibitem {Sheckelton_2017} Sheckelton J P, Plumb K W, Trump B A,Broholm C L, and McQueen T M 2017 {\it Inorg. Chem. Front.} {\bf 4} 481

	%\bibitem{Sheckelton_2017}
	%J. P. Sheckelton, K. W. Plumb, B. A. Trump, C. L. Broholm, and T. M. McQueen, Rearrangement of van der Waals stacking and formation of a singlet state at $T$ = 90 K in a cluster magnet, \href{https://doi.org/10.1039/C6QI00470A} {\emph{Inorg. Chem. Front.} {\bf 4}, 481 (2017).}

\bibitem {Jiang_2022} Jiang Z, Jiang D, Wang Y, Li C, Liu K, Wen T, Liu F, Zhou Z, and Wang Y 2022 {\it Sci. China Phys. Mech. Astron.} {\bf 65} 278211
	
	%\bibitem{Jiang_2022}
	%Z. Jiang, D. Jiang, Y. Wang, C. Li, K. Liu, T. Wen, F. Liu, Z. Zhou, and Y. Wang, Pressure-driven symmetry breaking and electron disproportionation of the trigonal Nb$_{3}$ cluster in Nb$_{3}$Cl$_{8}$, \href{https://doi.org/10.1007/s11433-022-1899-2} {\emph{Sci. China Phys. Mech. Astron.} {\bf 65}, 278211 (2022).}


\bibitem {Gao_2022} Gao S, Zhang S, Wang C, Tao W, Liu J, Wang T, Yuan S, Qu G, Pan M, Peng S, Hu Y, Li H, Huang Y, Zhou H, Meng S, Yang L, Wang Z, Yao Y, Chen Z, Shi M, Ding H, Jiang K, Li Y, Shi Y G, Weng H M, and Qian T 2023 {\it Phys. Rev. X} {\bf 13} 041049	

	%\bibitem{Gao_2022}
	%S. Gao, S. Zhang, C. Wang, W. Tao, J. Liu, T. Wang, S. Yuan, G. Qu, M. Pan, S. Peng, Y. Hu, H. Li, Y. Huang, H. Zhou, S. Meng, L. Yang, Z. Wang, Y. Yao, Z. Chen, M. Shi, H. Ding, K. Jiang, Y. Li, Y. Shi, H. Weng, and T. Qian, Discovery of a Single-Band Mott Insulator in a van der Waals Flat-Band Compound,
%\href{https://doi.org/10.1103/physrevx.13.041049}{\emph{Phys. Rev. X} \textbf{13}, 041049 (2023).}


\bibitem {Yoon_2020} Yoon J, Lesne E, Sklarek K, Sheckelton J, Pasco C, Parkin S S P, McQueen T M, and Ali M N 2020 {\it J. Phys.: Condens. Matter} {\bf 32} 304004	

	
%\bibitem{Yoon_2020}
	%J. Yoon, E. Lesne, K. Sklarek, J. Sheckelton, C. Pasco, S. S. P. Parkin, T. M. McQueen, and M. N. Ali, Anomalous thickness-dependent electrical conductivity in van der Waals layered transition metal halide, Nb$_ {3} $Cl$_ {8}$, \href{https://iopscience.iop.org/article/10.1088/1361-648X/ab832b} {\emph{J. Phys.: Condens. Matter} \textbf{32}, 304004 (2020).}


\bibitem {Sun_2022} Sun Z, Zhou H, Wang C, Kumar S, Geng D, Yue S, Han X, Haraguchi Y, Shimada K, Cheng P, Chen L, Shi Y, Wu K, Meng S, and Feng B 2022 {\it Nano Lett.} {\bf 22} 4596		
	
%\bibitem{Sun_2022}
	%Z. Sun, H. Zhou, C. Wang, S. Kumar, D. Geng, S. Yue, X. Han, Y. Haraguchi, K. Shimada, P. Cheng, L. Chen, Y. Shi, K. Wu, S. Meng, and B. Feng, Observation of Topological Flat Bands in the Kagome Semiconductor Nb$_ {3} $Cl$_ {8}$, \href{https://pubs.acs.org/doi/10.1021/acs.nanolett.2c00778} {\emph{Nano Lett. } \textbf{22}, 4596 (2022)}.	

\bibitem {Hu_2023} Hu J Y, Zhang X F, Hu C, Sun J, Wang X Q, Lin H Q, and Li G 2023 {\it Commun. Phys.} {\bf 6} 172
	
	%\bibitem{Hu_2023}
	%J. Y. Hu, X. F. Zhang, C. Hu, J. Sun, X. Q. Wang, H.-Q. Lin, and G. Li, Correlated flat bands and quantum spin liquid state in a cluster Mott insulator, \href{https://doi.org/10.1038/s42005-023-01292-z} {\emph{Commun. Phys.} {\bf 6}, 172 (2023).}


\bibitem {Grytsiuk_2024} Grytsiuk S, Katsnelson M I, van Loon E G C P, and R\"{o}sner M 2024 {\it npj Quantum Materials} {\bf 9} 8

%\bibitem{Grytsiuk_2024}
	%S. Grytsiuk, M. I. Katsnelson, E. G.C.P. van Loon, and M. R?sner, Nb$_ {3} $Cl$_ {8}$: a prototypical layered Mott-Hubbard insulator, \href{https://www.nature.com/articles/s41535-024-00619-5} {\emph{npj Quantum Materials} {\bf 9}, 8 (2024).}


\bibitem {Kasahara_2018} Kasahara Y, Ohnishi T, Mizukami Y, Tanaka O, Ma S, Sugii K, Kurita N, Tanaka H, Nasu J, Motome Y, Shibauchi T, and Matsuda Y 2018 {\it Nature} {\bf 559} 227	
	

	
	%\bibitem{Kasahara_2018}
	%Y. Kasahara, T. Ohnishi, Y. Mizukami, O. Tanaka, Sixiao Ma, K. Sugii, N. Kurita, H. Tanaka, J. Nasu, Y. Motome, T. Shibauchi, and Y. Matsuda, Majorana quantization and half-integer thermal quantum Hall effect in a Kitaev spin liquid, \href{https://doi.org/10.1038/s41586-018-0274-0} {\emph{Nature} {\bf 559}, 227 (2018)}.

\bibitem {Jiang_2017} Jiang J, Liang Q, Meng R, Yang Q, Tan C, Sun X, and Chen X 2017 {\it Nanoscale} {\bf 9} 2992

	
	%\bibitem{Jiang_2017}
	%J. Jiang, Q. Liang, R. Meng, Q. Yang, C. Tan, X. Sun, and X. Chen, Exploration of new ferromagnetic, semiconducting and biocompatible Nb$_ {3}X_ {8}$ ($X$ = Cl, Br or I) monolayers with considerable visible and infrared light absorption, \href{https://doi.org/10.1039/C6NR07231C} {\emph{Nanoscale} {\bf 9}, 2992 (2017)}.


\bibitem {Bradlyn_2017} Bradlyn B, Elcoro L, Cano J, Vergniory M G, Wang Z, Felser C, Aroyo M I, and Andrei Bernevig B 2017 {\it Nature} {\bf 547} 298
	
	%\bibitem{Bradlyn_2017}
	%B. Bradlyn, L. Elcoro, J. Cano, M. G. Vergniory, Z. Wang, C. Felser, M. I. Aroyo, and B. Andrei Bernevig, Topological quantum chemistry, \href{https://doi.org/10.1038/nature23268} {\emph{Nature} {\bf 547}, 298 (2017)}.

\bibitem {Zhang_2023} Zhang Y, Gu Y, Weng H, Jiang K, and Hu J 2023 {\it Phys. Rev. B} {\bf 107} 035126
	
	%\bibitem{Zhang_2023}
	%Y. Zhang, Y. Gu, H. Weng, K. Jiang, and J. Hu, Mottness in two-dimensional van der Waals Nb$_ {3}X_ {8}$ monolayers ($X$ =Cl, Br, and I), \href{https://link.aps.org/doi/10.1103/PhysRevB.107.035126} {\emph{Phys. Rev. B} {\bf 107}, 035126 (2023)}.

\bibitem {Wu_2022} Wu H, Wang Y, Xu Y, Sivakumar P K, Pasco C, Filippozzi U, Parkin S S P, Zeng Y  J, McQueen T, and Ali M N 2022 {\it Nature} {\bf 604} 653
	
	%\bibitem{Wu_2022}
	%H. Wu, Y. Wang, Y. Xu, P. K. Sivakumar, C. Pasco, U. Filippozzi, S. S. P. Parkin, Y.-J. Zeng, T. McQueen, and M. N. Ali, The field-free Josephson diode in a van der Waals heterostructure,  \href{https://doi.org/10.1038/s41586-022-04504-8}{\emph{Nature} {\bf 604}, 653(2022)}.


\bibitem {Ando_2020} Ando F, Miyasaka Y, Li T, Ishizuka J, Arakawa T, Shiota Y, Moriyama T, Yanase Y, and Ono T 2020 {\it Nature} {\bf 584} 373
	
	%\bibitem{Ando_2020}
	%F. Ando, Y. Miyasaka, T. Li, J. Ishizuka, T. Arakawa, Y. Shiota, T. Moriyama, Y. Yanase, and T. Ono, Observation of superconducting diode effect,   \href{https://doi.org/10.1038/s41586-020-2590-4}{\emph{Nature} {\bf 584}, 373 (2020)}.
	

\bibitem{SM} See Supplemental Materials for details.

\bibitem{takigawa} Takigawa M, Reyes A P, Hammel P C, Thompson J D, Heffner R H, Fisk Z, and Ott K C 1991 {\it Phys.
Rev. B} {\bf 43} 247
	
\bibitem {Moriya} Moriya T 1985 {\it Spin Fluctuations in Itinerant Electron Magnetism} (Springer-Verlag)

%\bibitem{Moriya}
%T. Moriya, Spin Fluctuations in Itinerant Electron Magnetism (Springer-Verlag, 1985).
\end{references}
\end{document}

\title{Supplymentary materials for Antiferromagnetic spin fluctuations and structural transition in cluster Mott insulator candidate Nb$_{3}$Cl$_{8}$ revealed by $^{93}$Nb- and $^{35}$Cl-NMR}

	\author{Y. Z. Zhou}
	\affiliation{Institute of Physics, Chinese Academy of Sciences,\\
		and Beijing National Laboratory for Condensed Matter Physics, Beijing 100190, China}
	\affiliation{School of Physical Sciences, University of Chinese Academy of Sciences, Beijing 100190, China}

   \author{X. Han}
	\affiliation{Institute of Physics, Chinese Academy of Sciences,\\
		and Beijing National Laboratory for Condensed Matter Physics, Beijing 100190, China}
	\affiliation{School of Physical Sciences, University of Chinese Academy of Sciences, Beijing 100190, China}

    \author{J. Luo}
	\affiliation{Institute of Physics, Chinese Academy of Sciences,\\
		and Beijing National Laboratory for Condensed Matter Physics, Beijing 100190, China}

    \author{D. T. Wu}
     \affiliation{Center for Neutron Science and Technology, Guangdong Provincial Key Laboratory of Magnetoelectric Physics and Devices, School of Physics, Sun Yat-Sen University, Guangzhou, Guangdong 510275, China}
     \affiliation{State Key Laboratory of Optoelectronic Materials and Technologies, Sun Yat-Sen University, Guangzhou, Guangdong 510275, China}
     \affiliation{Institute of Physics, Chinese Academy of Sciences,\\
		and Beijing National Laboratory for Condensed Matter Physics, Beijing 100190, China}

     \author{A. F. Fang}
     \affiliation{School of Physics and Astronomy, Beijing Normal University, Beijing 100875, China}
     \affiliation{Key Laboratory of Multiscale Spin Physics, Ministry of Education, Beijing Normal University, Beijing 100875, China}
     
    \author{B. Shen}
    \affiliation{Center for Neutron Science and Technology, Guangdong Provincial Key Laboratory of Magnetoelectric Physics and Devices, School of Physics, Sun Yat-Sen University, Guangzhou, Guangdong 510275, China}
    \affiliation{State Key Laboratory of Optoelectronic Materials and Technologies, Sun Yat-Sen University, Guangzhou, Guangdong 510275, China}
    
    \author{B. J. Feng}
	\affiliation{Institute of Physics, Chinese Academy of Sciences,\\
		and Beijing National Laboratory for Condensed Matter Physics, Beijing 100190, China}

    \author{Y. G. Shi}
    \email{ygshi@iphy.ac.cn}
	\affiliation{Institute of Physics, Chinese Academy of Sciences,\\
		and Beijing National Laboratory for Condensed Matter Physics, Beijing 100190, China}
	
    \author{J. Yang}
    \email{yangjie@iphy.ac.cn}
	\affiliation{Institute of Physics, Chinese Academy of Sciences,\\
		and Beijing National Laboratory for Condensed Matter Physics, Beijing 100190, China}

	\author{R. Zhou}
	\email{rzhou@iphy.ac.cn}
	\affiliation{Institute of Physics, Chinese Academy of Sciences,\\
		and Beijing National Laboratory for Condensed Matter Physics, Beijing 100190, China}

	\date{\today}% It is always \today, today,

	%\pacs{74.25.nj, 74.40.-n, 74.25.Dw}
	
	%\keywords{High-temperature superconductors, Nuclear magnetic resonance, pair density wave, Pseudogap}
	
	\maketitle

%	\section{$^{93}$Nb NMR spectrum at 10 K}
	
%	FIG.S1 shows the $^{93}$Nb NMR spectrum at 10 K by sweeping frequency with a fixed magnetic field $B_{0}$ = 4 T. The satellite peaks all split into two parts, with the relative abundance ratio of two Nb sits around 2:1. The relative abundance ratio, or to say stoichiometric ratio, which is determined by the ratio of the area of the splitting peaks, depicts the ratio of the quantity of nuclei of different sites. For each set of splitting peaks at 10 K and 78 K, the intervals are approximately equidistant, indicating the asymmetry parameter $\eta$ is close to 0. Here, $\eta = \dfrac{\lvert V_{xx}-V_{yy} \lvert}{V_{zz}}$, where $V_{\alpha\alpha}$ is the electric field gradient(EFG) of $\alpha=x, y, z$ directions. Table S1 shows the NMR parameters for both set of $^{93}$Nb peaks, including electric quadrupole frequency $\nu_{\rm Q}$, Knight shift $K$, asymmetry parameter $\eta$, and relative abundance ratio.

	\section{DC Magnetic Susceptibility}
	
	\begin{figure}[htbp]
		\includegraphics[width= 13 cm]{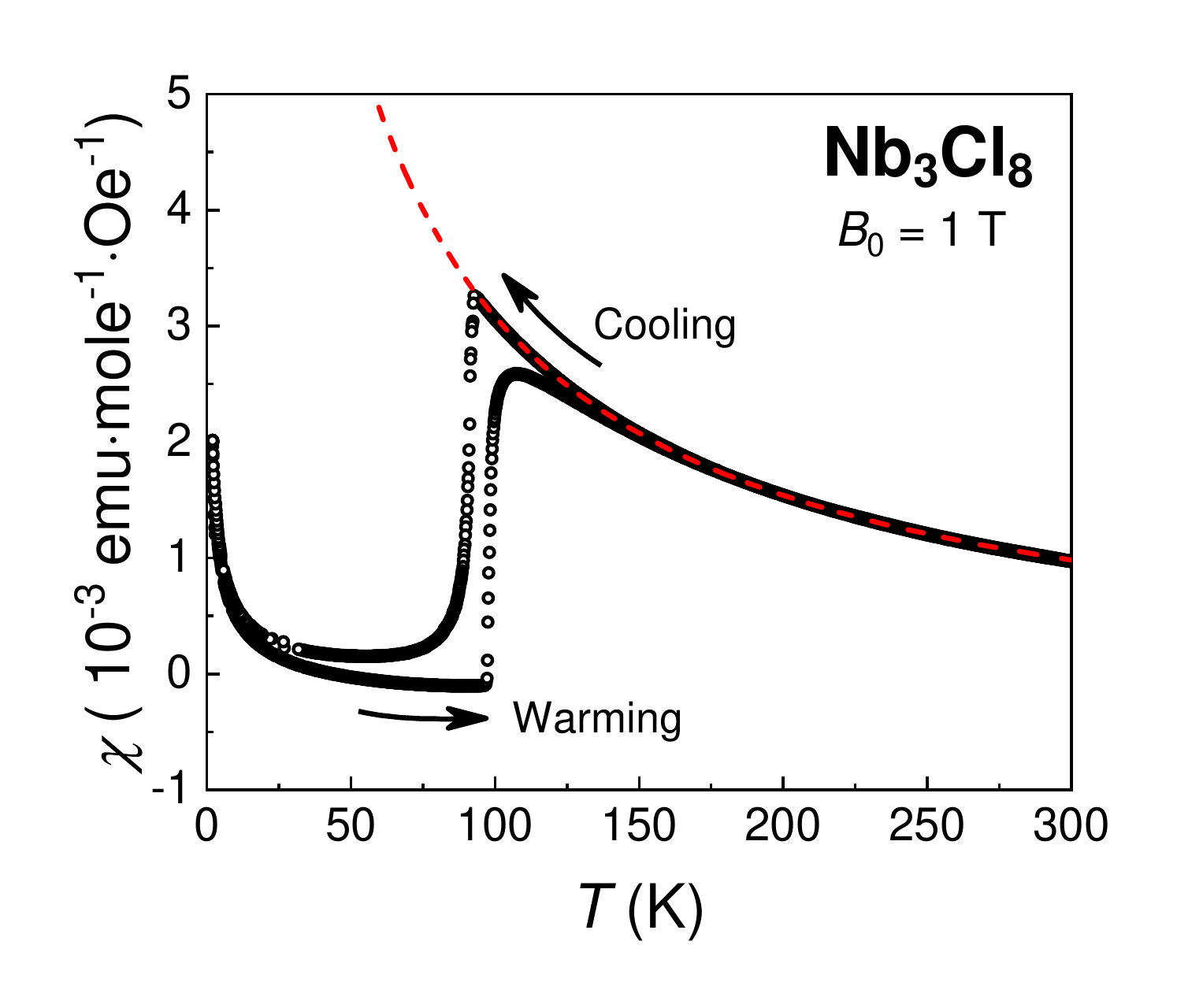}
		\renewcommand{\thefigure}{S\arabic{figure}}
		\caption{(Color online) DC magnetic susceptibility of Nb$_ {3} $Cl$_ {8}$ with applied field parallel to $c$-axis. The dashed line is a fit to the Curie-Weiss function for the high temperature phase.%, with Curie constant $C$= 0.3935 emu K mol$^{-1}$Oe$^{-1}$ and Weiss temperature $\theta$ = -18.4 K.
			\label{DC}}
	\end{figure}

	Figure \ref{DC} shows the temperature dependence of the DC magnetic susceptibility $\chi$ of a Nb$_ {3} $Cl$_ {8}$ single crystal sample with an applied field parallel to $c$-axis. A hysteretic phase transition occurs at around $T_{s}\sim$ 97 K. Below $T_{s}$, $\chi$ decreases rapidly to nearly zero, implying the low temperature phase is non-magnetic. The presence of a Curie tail at low temperatures is sample dependent and could result from paramagnetic impurities. In the  high temperature phase, $\chi$ can be well fitted by the Curie-Weiss law $\chi=\dfrac{C}{T-\theta}+\chi_{0}$, with Curie constant $C$= 0.3935 emu$\cdot$K$\cdot$mol$^{-1}$$\cdot$Oe$^{-1}$ and Weiss temperature $\theta$ = -18.4 K. The Curie constant $C$ can be converted into effective paramagnetic Bohr magneton $p_{\rm eff}$ by $C=p_{\rm eff}^{2}/8$. The obtained $p_{\rm eff}\approx$1.774 is very close to the theoretical value 1.73 for $S_{\rm eff}$=1/2, indicating that each Nb$_{3}$ cluster hosts a net unpaired spin on the molecular orbitals. Our magnetic susceptibility results are consistent with the previous report\cite{Haraguchi_2017}.

\clearpage

\section{Identification of the $^{35}$Cl-NMR peaks}

	\begin{figure}[htbp]
		\includegraphics[width= 12 cm]{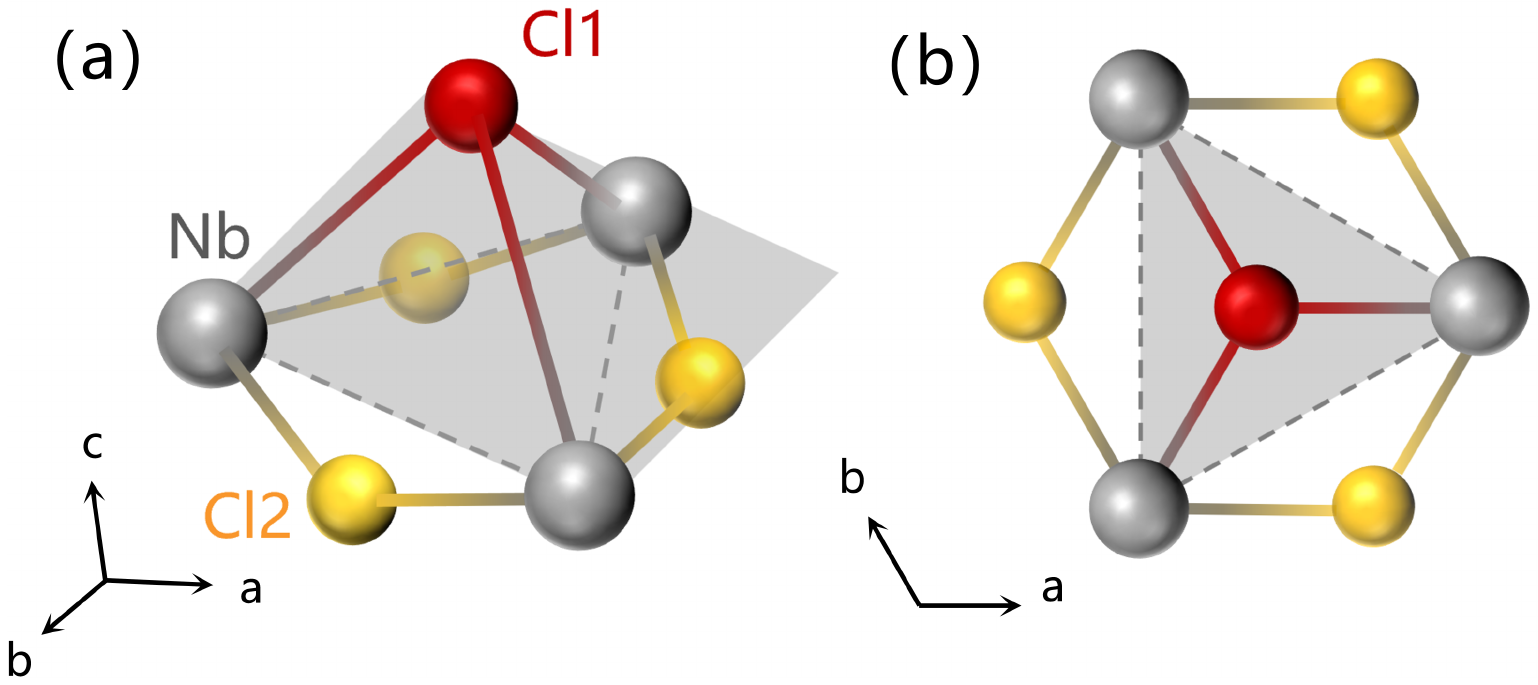}
		\renewcommand{\thefigure}{S\arabic{figure}}
		\caption{(Color online) The sketch for a Nb$_3$ triangle and surrounding Cl atoms. (a) Front view. (b) Top view.
			\label{lacal}}
	\end{figure}	

We elaborate the assignment of different Cl sites to the obtained $^{35}$Cl-NMR peaks. As can be seen from Fig. \ref{lacal}, Cl1 site is located right in the center of the Nb$_3$ triangle, where the EFG is symmetric. In contrast, Cl2 atoms are in  asymmetric EFG environment. Based on this fact, it is possible to distinguish Cl1 and Cl2 by their distinct EFG  asymmetry parameter $\eta$.  However, the $^{35}$Cl-NMR spectra we obtained are incomplete, so we can not deduce $\eta$ directly from the simulation of the spectrum to Eq. (1) in the main text.
When magnetic field is applied along the principal axis, that is $z$, the resonance frequency shift $\Delta f / f_{\rm L}$ is given by \cite{Abragam_1961}

\begin{figure}[htbp]
		\includegraphics[width= 12 cm]{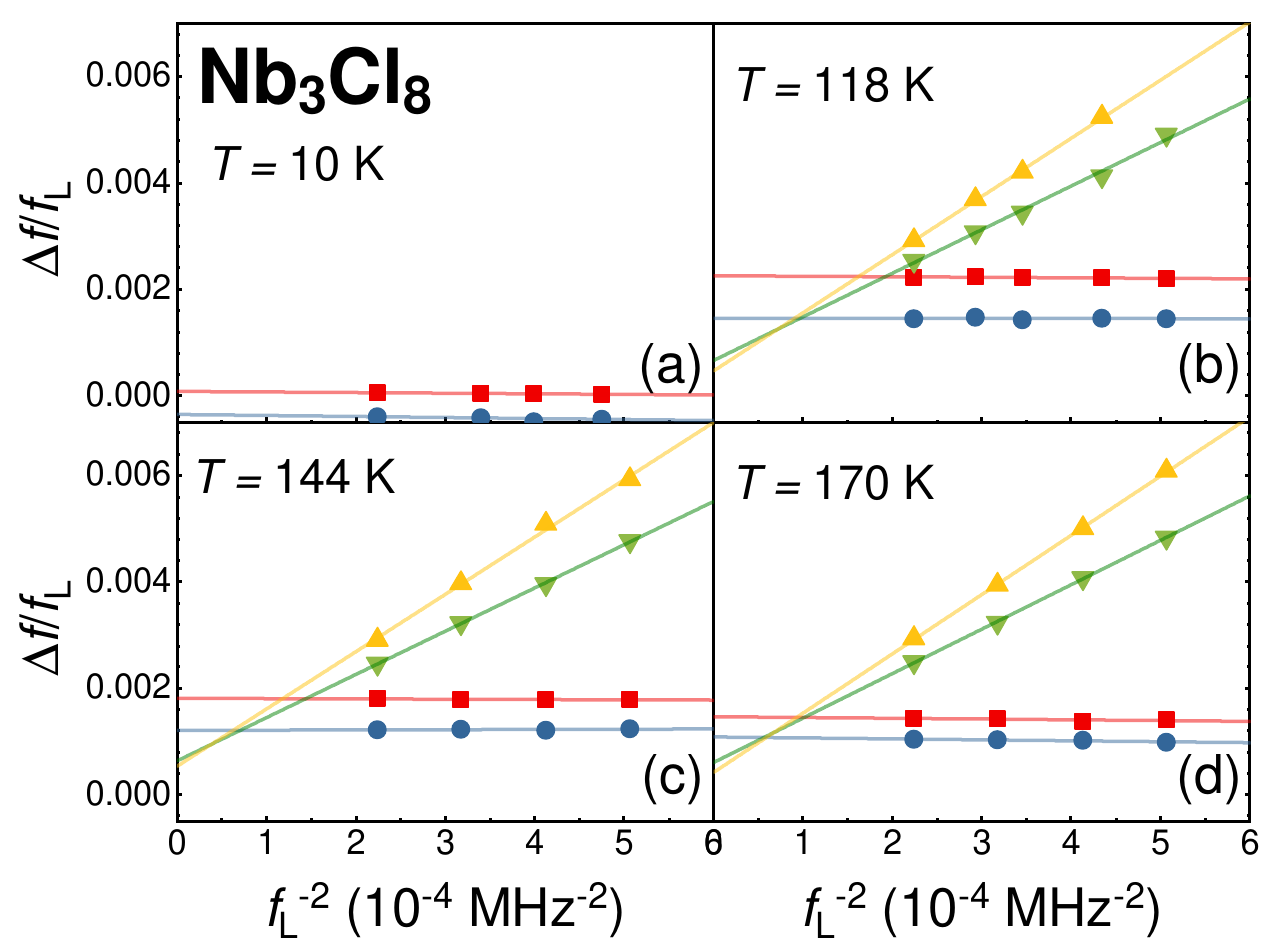}
		\renewcommand{\thefigure}{S\arabic{figure}}
		\caption{(Color online) The plot of $\Delta f / f_{\rm L}$ against $f_{\rm L}^{-2}$ for four $^{35}$Cl-NMR peaks at different temperatures. %(b)Temperture-dependence of $\dfrac{(v_{x}-v_{y})^2}{12(1 + K)}$which are obtained by the slope of $\Delta f \backslash f_{\rm L}$ versus $f_{\rm L}^{-2}$ lines.
			\label{Cl_ident}}
	\end{figure}

\begin{figure}[htbp]
		\includegraphics[width= 12 cm]{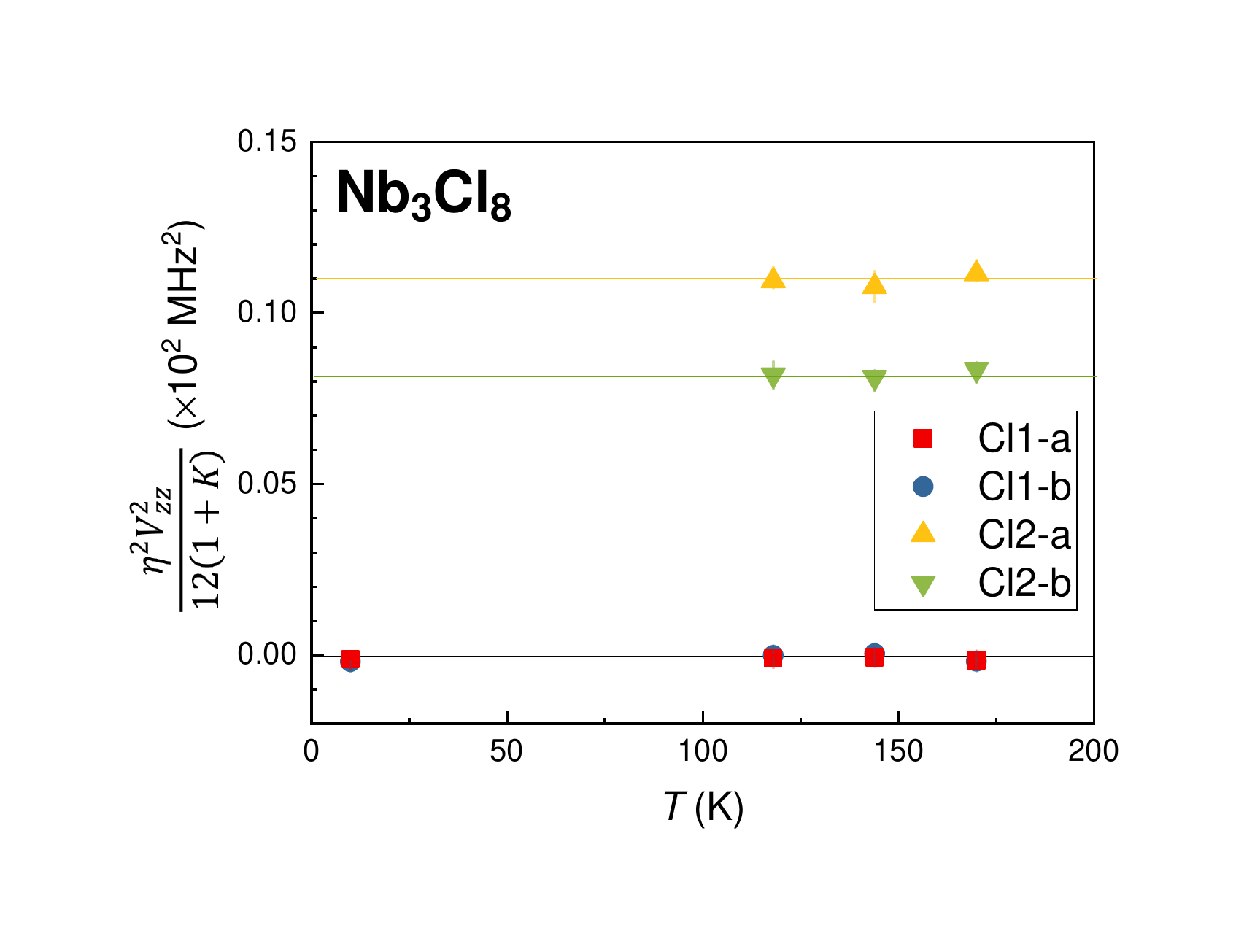}
		\renewcommand{\thefigure}{S\arabic{figure}}
		\caption{(Color online) The temperature dependence of $\dfrac{\eta^2\nu_{zz}^2}{12(1 + K)}$  obtained by the slope of $\Delta f / f_{\rm L}$ versus $f_{\rm L}^{-2}$ lines.
			\label{Cl_eta}}
	\end{figure}

% Due to different external environments for the four $^{35}$Cl sites in Nb$_{3}$Cl$_{13}$ unit cell, their response to external changes should have different trends by the numerical value and asymmetry of electric field gradient. Therefore, considering of the uncertain impact from EFG, a suitable approach here is to change the magnetic field along the principal axis, where the resonance frequency should follow as\cite{Abragam_1961}:

\begin{equation}
		\dfrac{\Delta f}{f_{\rm L}} = \dfrac{f_{\rm res} - f_{\rm L}}{f_{\rm L}} = K + \dfrac{(\nu_{x}-\nu_{y})^2}{12(1 + K)}\cdot\dfrac{1}{f_{\rm L}^{2}} = K + \dfrac{\eta^2\nu_{z}^2}{12(1 + K)}\cdot\dfrac{1}{f_{\rm L}^{2}}
\label{eqs1}
	\end{equation}
where $f_{res}$ is the observed resonance frequency, $f_L = \gamma B_0$ is the Lamor frequency and $\nu_{z}$ is the nuclear quadrupole resonance (NQR) frequency tensor. According to Eq. S\ref{eqs1}, $\eta$ can be obtained from the slope of the $\Delta f / f_{\rm L}$ against $f_{\rm L}^{-2}$ plot. Figure \ref{Cl_ident} shows such plots for the four $^{35}$Cl-NMR peaks at different temperatures. The slope, i.e. the quantity $\dfrac{\eta^2\nu_{zz}^2}{12(1 + K)}$ is shown in Fig. \ref{Cl_eta}.
The slopes from the same peak remain unchanged at different temperatures, indicating that the four peaks are originated from different $^{35}$Cl sites.
The two lower frequency peaks of $^{35}$Cl-NMR spectrum exhibites nearly zero $\eta$, which means that they origin from symmetric EFG environment and can be assigned to Cl1-a and Cl1-b. The two higher frequency peaks have finite values of $\eta$ and correspond to Cl2-a and Cl2-b sites.
In addition,  the observed two peaks below $T_{s}$ can be assigned to Cl1-a and C1-b sites since they have similar zero value slopes.
%the dependence with fields of four $^{35}$Cl NMR peaks at different temperatures are plotted as straight lines of $\Delta f / f_{\rm L}$ against $f_{\rm L}^{-2}$ in Fig.S2(a), where vividly present four separate linear trends. By extracting the linear slopes, we plot their relationship with temperature in Fig.S2(b). One can find the slopes from the same peak remain unchanged at different temperatures, thereby indicating that the four peaks are originated from different $^{35}$Cl sites. Considering the $\eta$ embodied in the slopes, the almost zero-value of peak $\rm \uppercase\expandafter{\romannumeral1}$ and $\rm \uppercase\expandafter{\romannumeral2}$ points to the their origin from isotropic sites, that is, the Cl1-a and C1-b sites in the cental position of three Nbs. While the finite value of peak $\rm \uppercase\expandafter{\romannumeral3}$ and $\rm \uppercase\expandafter{\romannumeral4}$ corresponds to anisotropic Cl2-a and Cl2-b sites. Besides, the observed two peaks below the phase transition temperature can be deduced to derivate from Cl1-a and C1-b sites due to the similar zero-value slopes. Note that, by this method, all the peaks in Fig.S2 are asserted to be central peaks with somehow no satellite peaks observesd.	

\clearpage

\section{$^{35}K-\chi$ plot}

	\begin{figure}[htbp]
		\includegraphics[width= 13 cm]{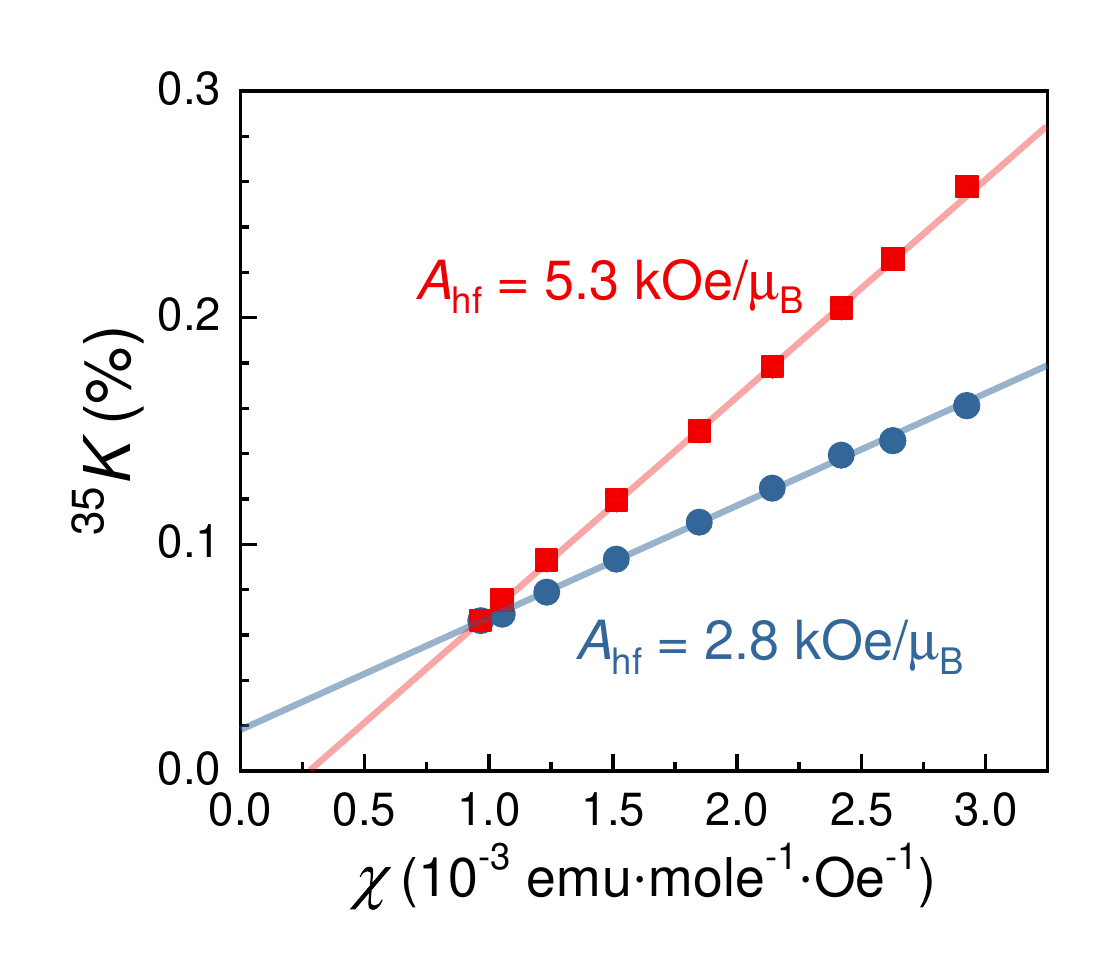}
		\renewcommand{\thefigure}{S\arabic{figure}}
		\caption{(Color online) $^{35}K-\chi$ plots for Cl1-a and Cl1-b sites, with temperature as an implicit parameter.
			\label{kchi}}
	\end{figure}

	Figure \ref{kchi} shows the Knight shift $K$ as a function of the DC magnetic susceptibility $\chi$ for Cl1-a and Cl1-b sites. Since $K_{\rm spin} = A_{\rm hf}\chi_{\rm spin}$, the hyperfine coupling constant $A_{\rm hf}$ can be extracted from the slope of the linear plot. $A_{\rm hf}$ = 5.3 and 2.8 kOe/$\mu_{\rm B}$  are obtained for the two Cl1 peaks. Considering that $A_{\rm hf}$ reflects the strength of the hyperfine interaction between electrons and nuclei, its value should be inversely proportional to the distance between $^{35}$Cl and Nb$_3$ trimer. Therefore, the NMR peak with $A_{\rm hf}$ = 5.3 kOe/$\mu_{\rm B}$ is assigned as corresponding to Cl1-a site, since the distance between Cl1-a and Nb$_3$ trimer is shorter compared to Cl1-b. The NMR peak with $A_{\rm hf}$ = 2.8 kOe/$\mu_{\rm B}$ is from the Cl1-b site.

\clearpage
\section{Hyperfine coupling constant}
	
\begin{figure}[htbp]
		\includegraphics[width= 14 cm]{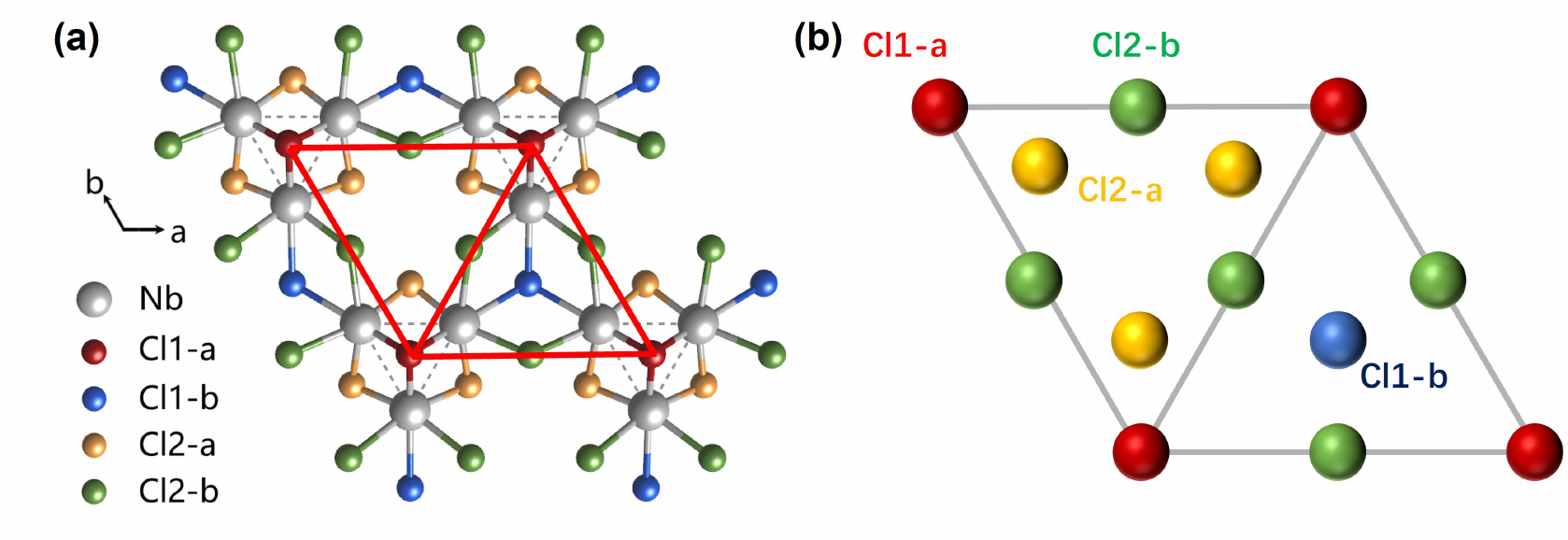}
		\renewcommand{\thefigure}{S\arabic{figure}}
		\caption{(Color online) (a)The basic structural unit of kagome lattice in Nb$_{3}$Cl$_{8}$ marked by a red parallelogram. (b) Relative positions of four Cl sites.
			\label{cluster}}
	\end{figure}

\begin{figure}[htbp]
		\includegraphics[width= 14 cm]{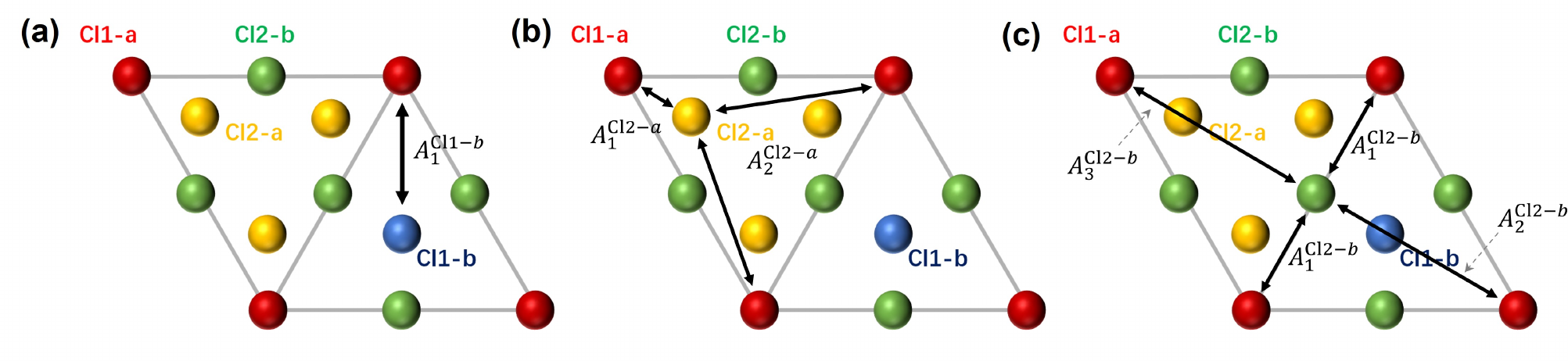}
		\renewcommand{\thefigure}{S\arabic{figure}}
		\caption{(Color online) The sketches of the hyperfine interactions between Cl nuclei and the nearest neighbor unpaired spins from the Nb$_{3}$ trimers. The Nb$_3$ trimer is  reduced as a point which is  just below Cl1-a.
			\label{hyperfine}}
	\end{figure}

%	FIG.S2(a) shows the temperature dependence of the DC magnetic susceptibility of a Nb$_ {3} $Cl$_ {8}$ single-crystal sample with an applied field parallel to $c$-axis. A hysteretic phase transition occurs at around $T^{*}\sim$100 K. Below $T^{*}$, the susceptibility $\chi$ rapidly decreases to nearly 0, implying the low-temperature phase is non-magnetic. The presence of a Curie tail at the lower temperature could result from paramagnetic impurity, which has been observed in 2D QSL candidate 1$T$-TaS$_{2}$. At high-temperature phase, the susceptibility conforms to the Curie-Weiss law $\chi=\dfrac{C}{T-\theta}+\chi_{0}$, with Curie constant $C$= 0.3935 emu K mol$^{-1}$ Oe$^{-1}$ and Weiss temperature $\theta$ = -18.4 K. Converting the Curie constant $C$ into effective paramagnetic Bohr magneton by $C=p_{\rm eff}^{2}/8$, the obtained $p_{\rm eff}\approx$1.774 approaches to the ideal value 1.73 for $S_{\rm eff}$=1/2, indicating that each Nb$_{3}$ cluster host a net unpaired spin on the molecular orbitals. Our susceptibility measurement is consistent with the previous results, which proves the sample quality.
According to $K_{\text{spin}} = A_{\text{spin}}\chi _{\text{spin}}$, the Knight shift $K$ is a local probe of the hyperfine interactions between the nuclear
spins and the surrounding electron spins. In order to analyze the surrounding environment of each Cl site, we attempt to simplify the crystal structure. The basic structural unit is shown in fig.~\ref{cluster}(a). The Nb$_3$ trimer can be  reduced as a point which is  just below Cl1-a, and then the basic unit of each layer's breathing kagome lattice can be reduced as a parallelogram structure.
Figure \ref{cluster}(b) shows the relative positions of each Cl site.
If we take into account only the hyperfine interactions between Cl atoms and the nearest neighbor Nb$_3$ trimers,  the hyperfine coupling constant $A_{\text{spin}}$ for each Cl sites can be estimated as follows.
For Cl1-a site, which is directly in the center of Nb$_3$ trimer, $A_{\text{spin}}^{\rm Cl1-a}$  will be determined by the unpaired spin of Nb$_3$ trimer. The Cl1-b site is influenced by three nearest Nb$_3$ trimers around it, as shown in fig.~\ref{hyperfine}(a). The  $A_{\text{spin}}^{\rm Cl1-b}$ can be expressed as

\begin{equation}
 A^{\mathrm{Cl1-b}}_{\mathrm{spin}}=3\cdot A^{\mathrm{Cl1-b}}_{1}
\end{equation}

Therefore, the temperature variation of $K$ for Cl1-a and Cl1-b sites will reflect the intrinsic properties
of electron spins, which is consist with the Curie-Weiss behavior of magnetic susceptibility.
As shown in fig.~\ref{hyperfine}(b) and (c), for the Cl2-a and Cl2-b sites, $A_{\text{spin}}$ can be expressed as

\begin{equation}
A^{\mathrm{Cl2-a}}_{\mathrm{spin}}=A^{\mathrm{Cl2-a}}_{1}+2\cdot A^{\mathrm{Cl2-a}}_{2}
\end{equation}
\begin{equation}
A^{\mathrm{Cl2-b}}_{\mathrm{spin}}=2\cdot A^{\mathrm{Cl2-b}}_{1}+A^{\mathrm{Cl2-b}}_{2}+A^{\mathrm{Cl2-b}}_{3}
\end{equation}

Due to the asymmetric environment, it is possible that $A_{\text{spin}}^{Cl2-a}$ and $A_{\text{spin}}^{Cl2-b}$ have a nearly zero value, resulting in temperature independent Knight shift for the Cl2-a and Cl2-b sites.
Similar results were reported in the cuprate YBa$_2$Cu$_3$O$_{6.63}$, in which $^{63}K_{c}$ is temperature independent but $^{63}K_{ab}$ changes with temperature varies\cite{Takigawa_1991}.
Through the analysis of  the hyperfine interactions between $^{63}$Cu and the nearest neighbor Cu 3$d$ spins, the hyperfine coupling constant happens to be nearly zero for the $c$-axis direction.

%$K^i_{\mathrm{spin}}\propto A^i_{\mathrm{spin}}\cdot\chi_{\mathrm{spin}}$

%
%\begin{acknowledgments}
%%	The authors thank  for valuable discussions.
%This work was supported by the National Natural Science Foundation of China (Grant Nos. 11974405 and Nos. 11904023), the National Key Research and Development Projects of China (Grant Nos. 2022YFA1403400) and the Strategic Priority Research Program of the Chinese Academy of Sciences (Grant Nos. XDB33010100 and Nos. XDB33030100).
%\end{acknowledgments}


%\vspace{0.5cm}
%\textbf{Author contributions}

%J. Luo and Z. Zhao contributed equally to this work. The single crystals were grown by Z.Z., H.T.Y. and H.J.G. The NMR measurements were performed by J.L., Z.Y.Z., J.Y., A.F.F. and R.Z. R.Z. and G.-q.Z. wrote the manuscript with inputs from J.L. All authors have discussed the results and the interpretation.

%\vspace{0.5cm}

%\textbf{Competing Interests}

%The authors declare no competing interests.